\documentclass[12pt,preprint]{aastex}

\shorttitle{Radial dependence of the mass function of IC~2391}
\shortauthors{Boudreault \& Bailer-Jones}

\begin{document}


\title{A constraint on brown dwarf formation via ejection: radial
  variation of the stellar and substellar mass function of the young
  open cluster IC~2391}


\author{S. Boudreault and C. A. L. Bailer-Jones}
 \affil{Max-Planck-Institut f\"ur Astronomie, K\"onigstuhl 17,
  Heidelberg, GERMANY 69117}




\begin{abstract}
  We present the stellar and substellar mass function of the open
  cluster IC~2391, plus its radial dependence, and use this to put
  constraints on the formation mechanism of brown dwarfs. Our
  multiband optical and infrared photometric survey with spectroscopic
  follow-up covers 11 square degrees, making it the largest survey of
  this cluster to date. We observe a radial variation in the mass
  function over the range 0.072 to 0.3\,M$_\odot$, but no significant
  variation in the mass function below the substellar boundary at the
  three cluster radius intervals analyzed.  This lack of radial
  variation for low masses is what we would expect with the ejection
  scenario for brown dwarf formation, although considering that
  IC~2391 has an age about three times older than its crossing time,
  we expect that brown dwarfs with a velocity greater than the escape
  velocity have already escaped the cluster. Alternatively, the
  variation in the mass function of the stellar objects could be an
  indication that they have undergone mass segregation via dynamical
  evolution. We also observe a significant variation across the
  cluster in the colour of the (background) field star locus in
  colour-magnitude diagrams and conclude that this is due to variable
  background extinction in the Galactic plane. From our preliminary
  spectroscopic follow-up to confirm brown dwarf status and cluster
  membership, we find that all candidates are M dwarfs (in either the
  field or the cluster), demonstrating the efficiency of our
  photometric selection method in avoiding contaminants (e.g.\ red
  giants).  About half of our photometric candidates for which we have
  spectra are spectroscopically-confirmed as cluster members; two are
  new spectroscopically-confirmed brown dwarf members of IC~2391.
\end{abstract}





\section{\label{intro} INTRODUCTION}

The origin and evolution of brown dwarfs (BD) remains a fundamental
open question.  BDs have masses bridging the lowest mass
hydrogen-burning stars and giant planets, so any picture of star and
planet formation is incomplete if it cannot account for BDs.  Several
formation mechanisms have been proposed, including star-like formation
from the compression and fragmentation of a dense molecular cloud,
planet-like formation in a circumstellar disk, and the dynamical
interruption of a star-like accretion process.

There are observational signatures which may be used to distinguish
between these scenarios, such as the distribution of binaries, the
presence and properties of circumstellar disks, the (initial) mass
function (MF) and kinematics (see \citealt{luhman2007} for a review of
observational signatures on the formation of BDs).  Work over the past
ten years has seen considerable success in measuring the MF into the
BD regime in several clusters, including $\sigma$~Orionis
(\citealt{gonsalez-garcia2006}; \citealt{caballero2007};
\citealt{lodieu2009}), the Orion Nebula Cluster (ONC)
(\citealt{hillenbrand2000}; \citealt{slesnick2004}), IC~2391
(\citealt{barrado2004}; \citealt{platais2007}) and the Pleiades
(\citealt{moraux2003}; \citealt{lodieu2007}). These are only examples
among many other analysis of stellar and substellar populations.

The comparison of the MF in clusters with different properties (e.g.\
the different density clusters Taurus and ONC; clusters with different
ages, e.g.\ \citealt{chabrier2003}) has led some workers to draw
conclusions about the relative efficiency of possible BD formation
mechanisms (e.g.\ \citealt{kroupa2003}). While some observations
(\citealt{luhman2007}) and theoretical works (\citealt{padoan2004};
\citealt{hennebelle2008}) conclude in a common formation mechanism for
BDs and stars, some studies has suggested that BDs could form by
massive disc fragmentation (\citealt{stamatellos2008}),
photoevaporation of the accretion envelope (\citealt{hester1996}), or
interruption of the accretion process (\citealt{reipurth2000};
\citealt{reipurth2001}). For instance, \cite{bate2005} have performed
hydrodynamical simulations of star formation from fragmentation of
molecular clouds.  They concluded that objects which end up as BDs
stop accreting before they reach the hydrogen burning limit because
they are ejected from the dense gas soon after their formation by
dynamical interaction in unstable multiple systems.

This ejection scenario in some cases predicts a higher velocity
dispersion and spatial spread of BDs in comparison to stellar objects,
which in turn may be observed as a variation in the MF with radius
(\citealt{kroupa2003}). On the other hand, other work have shown that
if BDs are formed by ejection, the velocity distribution could be the
same for BDs and stars (\citealt{bate2009}). \cite{muench2003}
observed a radial variation in the MF of IC~348 measured over 0.5 to
0.08\,M$_{\odot}$, but no variation was detected in the BD regime. In
a study of the spatial distribution of substellar objects in IC~348 as
well as Trapezium in the Orion Nebula Cluster, \cite{kumar2007}
observed the stellar objects to be more clustered than the substellar
ones, which they took as evidence in favour of the ejection scenario.
By looking at the spatial distribution of the Taurus stellar and
substellar population, \cite{guieu2006} observed a gradient in the BD
abundance relative to stars, which they conclude as in favour of the
ejection scenario.

In this paper, we present the results of a program to study, in
detail, the MF of one of the nearest and richest open clusters,
IC~2391. This has an age of 50\,Myr measured from lithium depletion
(\citealt{barrado2004}) or 40\,Myr from main-sequence fitting
(\citealt{platais2007}). The Hipparcos distance is
146.0$^{+4.8}_{-4.5}$\,pc (\citealt{robichon1999}) and the metalicity
and extinction are [Fe/H]$=-0.03 \pm 0.07$ and $E(B-V)=0.01$,
(\citealt{randich2001}).  Because of its proximity and youth, this
cluster has been subject to several studies (e.g.\
\citealt{barrado2004,barrado2001,barrado1999}; \citealt{koen2006};
\citealt{siegler2007}; \citealt{platais2007}).  \cite{barrado2004} and
\cite{dodd2004} measured the MF down to the substellar limit, but the
MF for confirmed member with a completeness limit at lower masses has
not yet been determined.

Since IC~2391 is not as young as IC~348 (age\,$\sim$\,2\,Myr from
\citealt{muench2003}) and Trapezium (age\,$\sim$\,0.8\,Myr from
\citealt{muench2002}), one may expect it has already lost a
significant fraction of its substellar population to evaporation by
dynamical evolution.  Using the tidal radius and mass of IC~2391
estimated by \cite{piskunov2007} (7.4\,pc and 175\,M$_{\odot}$), we
compute the escape velocity as $v_e=$0.4\,km/s and the crossing as
time $t_{cross}=$17\,Myr. (We stress that the crossing time is just an
order-of-magnitude quantity, 2R/v, where we have adopted for R the
tidal radius of 7.4\,pc from \citealt{piskunov2007} and the velocity
dispersion of $\sim$0.85\,km/s from \citealt{platais2007}).  Assuming
a minimum value for the number of cluster members as the objects
reported by \cite{dodd2004} and \cite{barrado2004} (125 and 33
respectively, together a total lower limit of 158 objects), we
estimate that the lower limit of the relaxation time for this cluster
is $t_{relax}\sim$105\,Myr.  Furthermore, in a numerical simulation of
open clusters and the population of BDs members, \cite{adams2002}
shows that there would still be more than 80\% of the original BD
population in the cluster even after about 10 crossing times, assuming
that BDs and stars have a similar velocity dispersion.  Therefore,
IC~2391 is still young enough for a radial study of its very low mass
star (VLMS) and BD populations, considering the fact that mass
segregation occurs on a timescale of order one relaxation time
(\citealt{bonnell1998}; although recent work by \citealt{allison2009}
suggest that mass segregation can occur on a smaller time than a
relaxation time, at least for more massive stars).

The paper is structured as follows. We will first present the data
set, reduction procedure and calibration in \S \ref{obs-data-calib}.
We then discuss our candidate selection procedure in \S
\ref{selection}, present the survey results in \S \ref{results-survey}
and then discuss the radial variation of the MF in \S
\ref{mf-variation}. The preliminary spectroscopic follow-up is
presented in \S \ref{spec-follow-up} followed by our conclusions in \S
\ref{conclusion}.

\section{\label{obs-data-calib} OBSERVATIONS, DATA REDUCTIONS AND
  CALIBRATIONS}

\subsection{\label{obs} Observations}

\begin{figure}
  \epsscale{1.0} \plotone{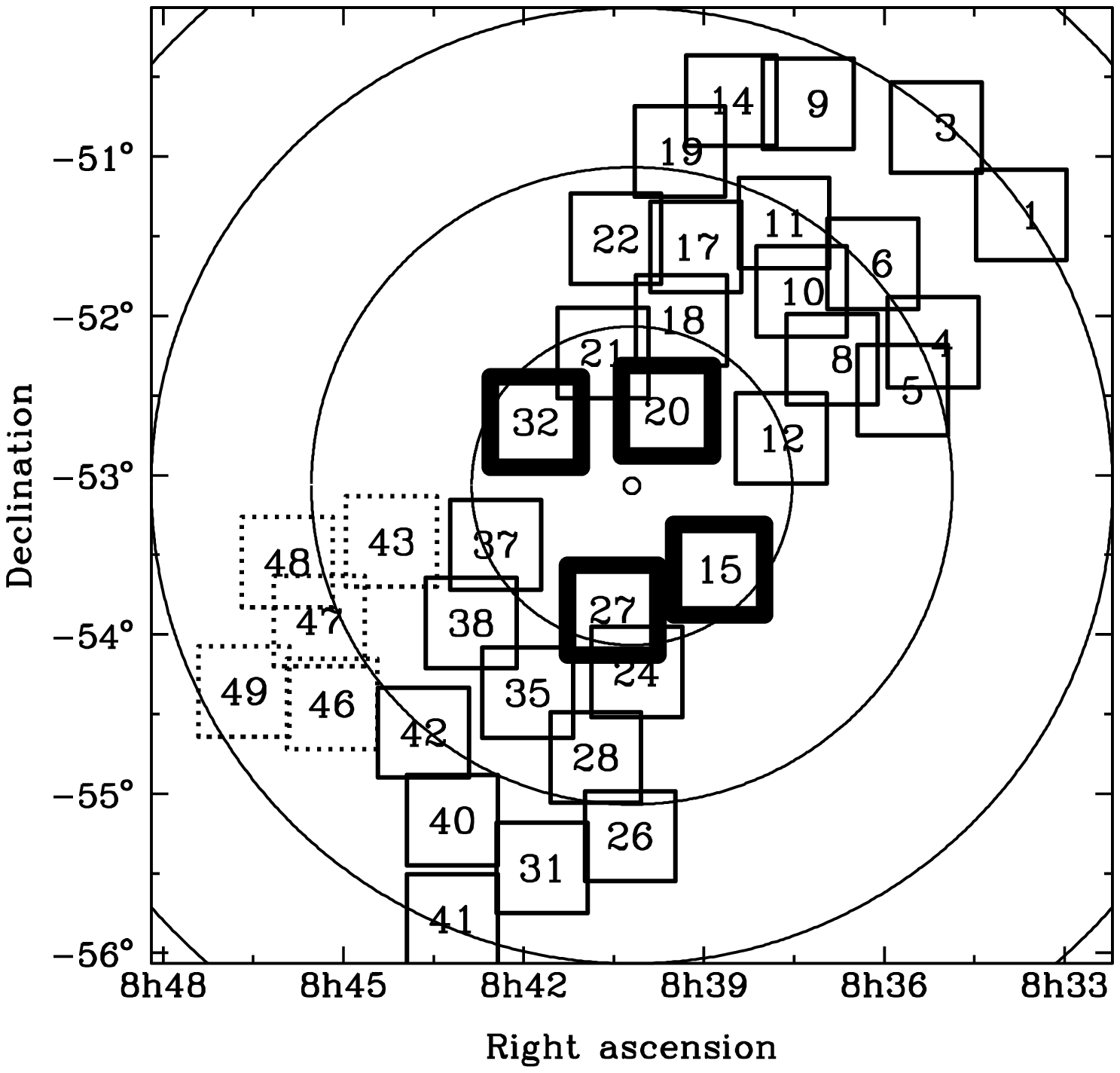}
  \caption{\label{fig:ic2391map} Area covered by our survey of IC
    2391. The four thick squares are the deep fields, the dotted
    squares are the outward fields and the others are the radial
    fields.  The circles have radii of 1$^\circ$, 2$^\circ$ and
    3$^\circ$ from the cluster center.}
\end{figure}

The survey consists of 35 34$\times$33 arcmin fields extending to 3
degrees from the center of the cluster and centered on RA=08:40:36
DEC=-53:02:00 (Figure \ref{fig:ic2391map}). The central 4 fields will
be referred to as the \textit{deep fields} while the other 31 other
fields will be referred as the \textit{radial fields} and
\textit{outward fields}. (We make a distinction since they were
observed with different exposure times and different filters, as will
be specified below). The fields were chosen to extend preferentially
along lines of constant Galactic latitude in an attempt to reduce
systematic errors in any established cluster MF gradient which could
arise from contamination by a Galactic disk population gradient. The
total coverage of our survey is 10.9 sq.\ deg.  This compares to 2.5
deg$^2$ in the survey of \cite{barrado2001}.

The optical observations were carried out in four runs with the Wide
Field Image (WFI) on the 2.2m telescope at La Silla
(\citealt{baade1999}) in : 24 January - 9 February 1999, 20 - 24
January 2000, 10 - 23 March 2007 and 15 - 18 May 2007.  The WFI is a
mosaic camera comprising 4$\times$2 CCDs each with 2k$\times$4k pixels
delivering a total field of view of 34$\times$33 arcmin at 0.238
arcsec per pixel.  The deep fields in our survey were observed in four
medium bands filters, namely 770/19, 815/20, 856/14 and 914/27 (where
the filter name notation is central wavelength on the full width at
half maximum, FWHM, in nm) and one broad band filter, $R_{\rm c}$.
The radial fields were observed in $R_{\rm c}$, 815/20 and 914/27
while the five outward fields were not observed in $R_{\rm c}$.

\begin{figure}
  \epsscale{1.0} \plotone{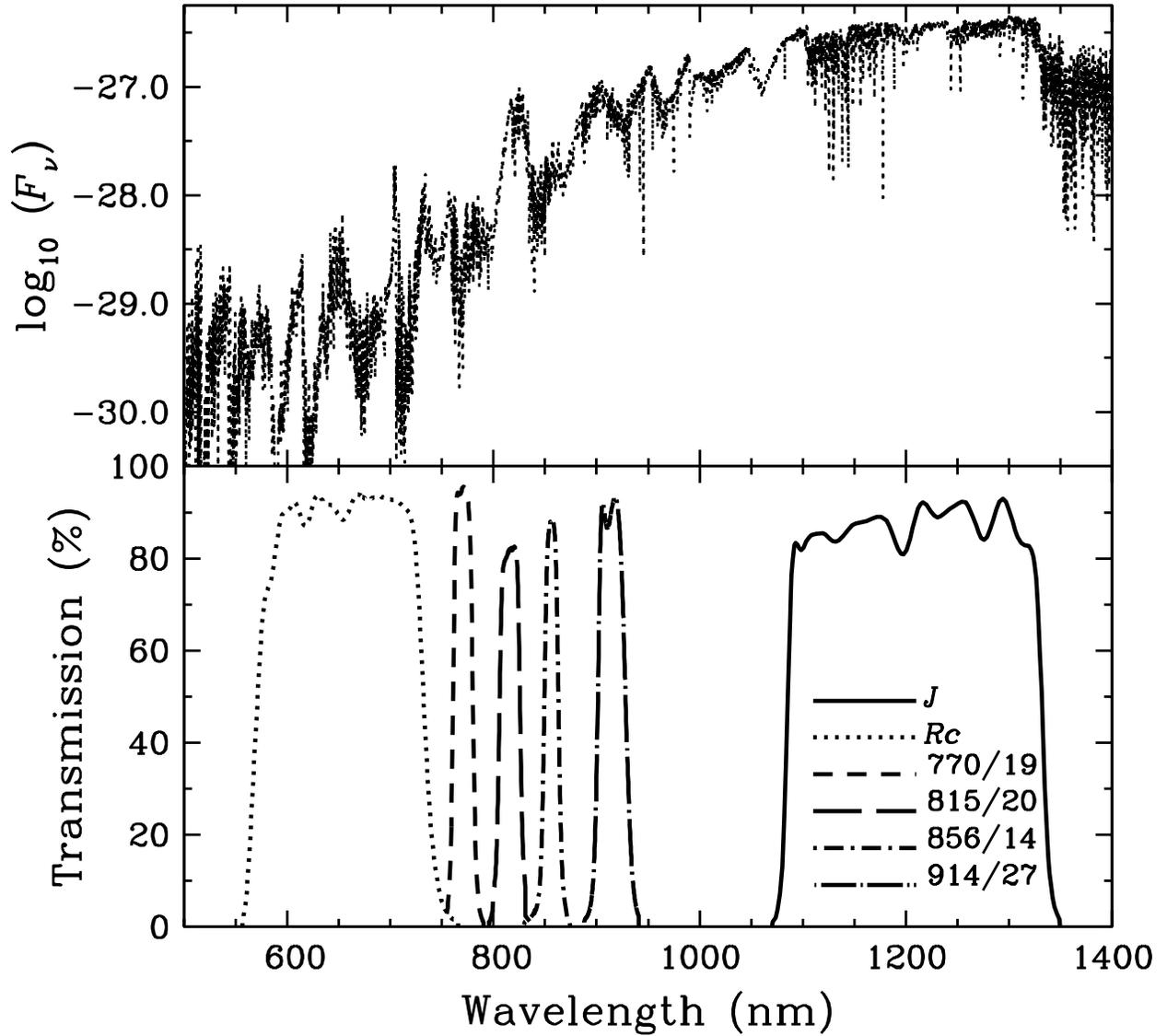}
  \caption{\label{fig:passband} Transmission curve of the filters used
    in our survey compared to the synthetic spectrum of a brown dwarf
    with $T_{\rm eff}$ = 2300 K, log \textit{g} = 4.5 and solar
    metallicity (NextGen model).}
\end{figure}

These filters were chosen to sample the spectra of late M and early L
dwarfs to improve selection over, say, $R_{\rm c}-I_{\rm c}$, and to
minimize the Earth-sky background plus any nebular emission (as the
filters are in regions of low emission). The pass band function for
all filters are shown in Figure \ref{fig:passband}. For all radial
fields, we have used an exposure time of 15, 10 and 25 minutes for the
815/20, 914/27 and $R_{\rm c}$ filters respectively. For each deep
field we obtained an integration time of 65 minutes for each of 770/19
and $R_{\rm c}$ and from 50 to 155 minutes for 815/20, from 25 to 261
minutes for 856/14, and 50 to 80 minutes for 914/27. We additionally
obtained short exposures for all fields to extend the dynamic range to
brighter objects. The photometry from the short exposures were
combined with the photometry of our long exposures for our analysis.
In order to improve the determination of their low mass status (via a
better determination of spectral type and luminosity), we also
observed all radial fields, including the outward fields, in the
$J$--band using the \textit{Cam\'era PAnoramique Proche-InfraRouge}
(CPAPIR) on the 1.5m telescope at Cerro Tololo, Chile (runs on 28
February - 3 March 2007 and 10 March 2007). However, we did not get
$J$--band photometry for the deep fields.  CPAPIR consists of one
Hawaii II detector of 2k$\times$2k pixels for a field of view of
35$\times$35 arcmin with a pixel scale of 1.03 arcsec per pixel. All
fields were observed with a total exposure time of 30 minutes. The $J$
filter of CPAPIR is centered at 1~250 nm with a FWHM of 160 nm. A
detailed list of the fields observed with pointing, filter used,
exposure time and 10$\sigma$ detection limit, is given in Table
\ref{tab:observations}.

\begin{figure}
  \epsscale{1.0} \plottwo{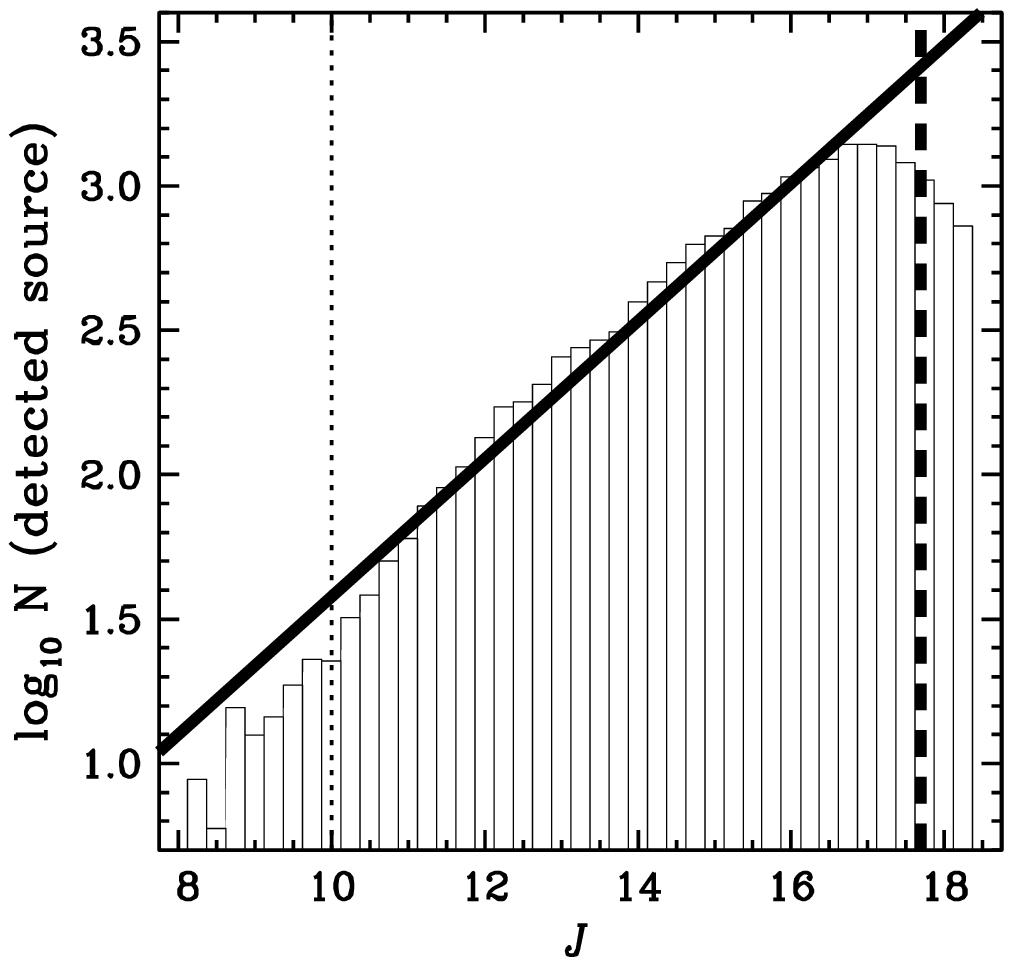}{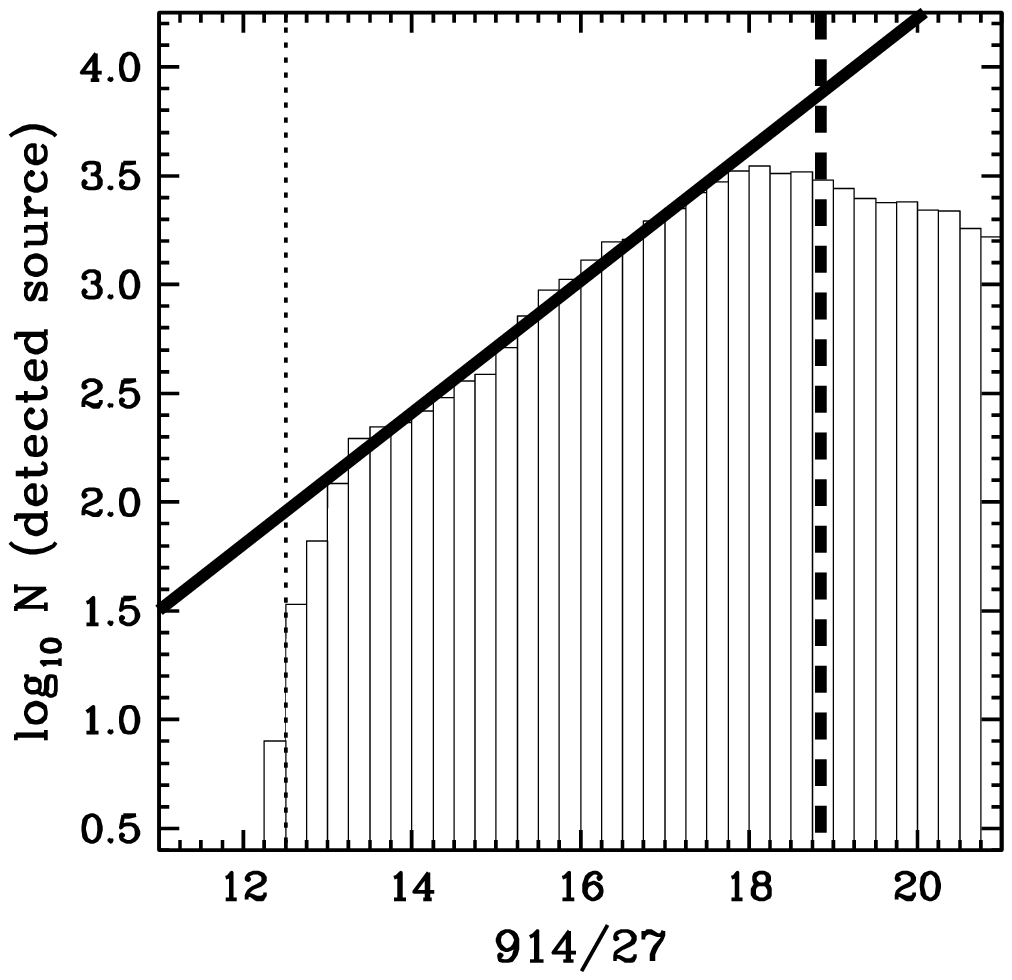}
  \caption{\label{fig:completeness} Estimation of the completeness
    limit for the radial part of our survey using the $J$ band
    (\textit{left}) and of the deep part using 914/27
    (\textit{right}).  The tick lines give best linear fit before
    the turn off; the vertical tick dotted line is the 10\,$\sigma$
    detection limit and the vertical thin line is the magnitude for
    which saturation start to occur in the short exposures.}
\end{figure}

Our 10$\sigma$ detection limit is $J$=17.7 and $914/27$=20.5 for the
radial and deep fields respectively (which corresponds to
$\sim$0.03\,M$_\odot$ for both cases). However, we can't expect to
detect \textit{all} objects down to these magnitudes.  The
completeness spanning from the brightest objects without saturation
down to the 10$\sigma$ detection limit is estimated by taking the
ratio of the number of objects detected to the predicted number of
detections and assuming a uniform distribution of stars along the line
of sight in the fields of our survey. The predicted number of
detections is derived very simply by extrapolating to the detection
limit a linear fit to the histogram of the number of detections as a
function of magnitude (Figure \ref{fig:completeness}).  The
completeness of the radial part of the survey is 91.8\% while for the
deep part it is 82.7\%.

The spectroscopic observations were carried out with HYDRA, a
multi-object, fiber-fed spectrograph on the 4m telescope at Cerro
Tololo on the nights of 6 and 7 January 2007. Only two fields could be
observed: the deep fields 15 and 20 (3 and 2 exposures of 45 minutes
respectively). We used the red fiber cable with the KPGLF grating (632
lines mm$^{-1}$) and a blaze angle of 14.7$^{\circ}$ (no blocking
filter was used) and . This gives us a coverage of 6429--8760\,\AA\,
centered at 7593\,\AA\, and a spectral resolution of 4.0\,\AA.

\subsection{\label{data} Reduction and Astrometry}

The standard CCD reduction steps (overscan subtraction, trimming and
flat-fielding for the WFI data and dark subtraction, flat-fielding for
CPAPIR data) were performed on a nightly basis using the $ccdred$
package under IRAF. For WFI data we used the dome flat both for
pixel-to-pixel variation correction and to correct the global
illumination, while for CPAPIR data we used \textit{superflat}
(obtained by combining science image frames for each nights).  For WFI
data, we reduced each of the 8 CCDs in the mosaic independently and in
the final step scaled them to a common flux response level.  We made
an initial sky subtraction via a low-order fit to the optical data,
and for the infrared data by subtracting a median combination of all
(unregistered) images of the science frames.  Images were fringe
subtracted when fringes were visible, which was the case for all
medium bands filters used, in a similar way as described by
\cite{calj2001}\footnote{A \textit{fringe correction frame} was
  created, which is a median combination of all science in a same
  filter with same exposure time. This \textit{fringe correction
    frame} was scaled by a factor -- determined manually for each
  science frames -- and subtracted from the science image.}  Finally,
the individual images of a given field were registered and median
combined.  We calculated magnitudes via aperture photometry together
with an aperture correction following the technique used in
\cite{howell1989}. An astrometric solution was achieved using the IRAF
package \textit{imcoords} and the tasks \textit{ccxymatch},
\textit{ccmap} and \textit{cctran}. For each field, this solution was
computed for the 815/20 band image using the \textit{Two Micron All
  Sky Survey} (2MASS) catalogue as a reference.  The RMS accuracy of
our astrometric solution is within 0.15--0.20 arcsec for WFI data and
within 0.3--0.4 arcsec for CPAPIR data.  For WFI data, the astrometry
was also performed on a CCD-by-CCD basis.

We only retained images for this study which were taken under
photometric conditions, as determined by our monitoring of conditions
during the observations and, moreover, by our data reduction
procedure.

\subsection{\label{calib} Photometric Calibration}

To correct for Earth-atmospheric absorption on the photometry, we
solved by least squares fitting the equation,
\begin{equation}
  {m_{A}=m_{A,raw}+Z_A+C_A(m_{A}-m_{B})+\kappa_AX_A},
\label{eqw:phot-cal}
\end{equation}
\noindent for observations of the standard stars at a range of
airmasses, where the spectrophotometric standard stars observed were
Hiltner~600, HR~3454 and LTT~3864. (In order to obtain the observed
magnitude from equations \ref{eqw:phot-cal}, the fluxes of each
standard star, $f_{\lambda}$, were taken from
\citealt{hamuy1992,hamuy1994}.) The parameters $m_{A}$ and $m_{B}$ are
the apparent magnitudes of our spectrophotometric standard in two
particular bands ($A$ and $B$), where $m_{A,raw}$ is the instrumental
magnitude of our spectrophotometric standard stars, $Z_A$ is the zero
point offset, $C_A$ is the colour correction and $\kappa_A$ the
extinction coefficient for band $A$ and $X$ is the airmass at which
$m_{A,raw}$ was obtained.

We calibrated the infrared data using the $J$ band values of 2MASS
objects which were observed in the science fields. By determining a
constant offset between the magnitude of 2MASS and our instrumental
magnitude, we obtained the zero point offset. Since this zero point
offset was obtained with objects in the same field of view in each
science frame, we did not perform a colour or airmass correction when
reducing our NIR photometry.

\subsection{\label{get-mass} Mass and Effective Temperature Based on
  Photometry}

We used the spectral energy distribution to derive the mass and
effective temperature, $T_{\rm eff}$, assuming that all our
photometric candidates belong to IC~2391. We used evolutionary tracks
from \cite{baraffe1998} and atmosphere models from
\cite{hauschildt1999a} (assuming a dust-free atmosphere; the NextGen
model) to compute an isochrone for IC~2391 using an age of 50\,Myr,
distance of 146\,pc, a solar metalicity and neglecting the reddening
($E(B-V)$\,=\,0). These models and assumptions provide us with a
prediction of $f_{\lambda}$, the spectral energy distribution received
at the Earth (in erg cm$^{-2}$ s$^{-1}$ \AA$^{-1}$) from the source.
We need to convert these spectra to magnitudes in the filters we used.
Denoting as $S_{A}$($\lambda$) the (known) total transmission function
of filter $A$ (including the CCD quantum efficiency and assuming
telescope and instrumental throughput is flat), then the flux measured
in the filter is

\begin{equation}
  f_{A}=\frac{\int_{0}^{\infty}f_{\lambda}S_{A}(\lambda)d\lambda}{\int_{0}^{\infty}S_{A}(\lambda)d\lambda},
  \label{eqw:phot-model}
\end{equation}

\noindent The corresponding magnitude $m_{A}$ in the Johnson
photometric system is given by

\begin{equation} {m_{A}=-2.5\;\mathrm{log}\;f_{A}\;+\;c_{A}},
  \label{eqw:phot-cal-2}
\end{equation}

\noindent where $c_{A}$ is a constant (zero point) that remains to be
determined in order to put the model-predicted magnitude onto the
Johnson system.  We determine this constant for each of the bands
$R_{\rm c}$, 770/19, 815/20, 856/14, 914/27 and $J$ in the standard
way, namely by requiring that the spectrum of Vega produce a
magnitude, $m_{A}$, of zero in all bands. Using the Vega spectrum from
\cite{colina1996} we derive values of $c_{Rc}=-21.6607$
$c_{770/19}=-22.2517$, $c_{815/20}=-22.4391$, $c_{856/14}=-22.6341$,
$c_{914/27}=-22.8353$ and $c_{J}=-23.6865$\,mag. Applying the two
equations above to a whole set of model spectra produces a theoretical
isochrone in colour--magnitude space.  Note that this procedure only
provides us with the ``true'' magnitudes of the model spectra, not
their instrumental ones. The photometric calibration procedure applied
to the data converts the measured, instrumental magnitudes to the
``true'' magnitude plane where we can compare them.

Assuming that all our photometric candidates belong to IC~2391, we
derive masses and effective temperatures in the following way.  We
first normalize the measured spectral energy distribution (multiband
photometry) of each object to the energy distribution of the model
using the 815/20 filter.  We then estimate the mass and effective
temperature (which are not independent of course) via a least squares
fit between the measured spectral energy distribution and the model
spectral energy distribution from the isochrone.

There are several sources of error in the mass and $T_{\rm eff}$
estimates. These come from the photometry, the photometric
calibration, the least squares fitting (imperfect model) and the
uncertainties on the age of IC~2391 (we use 5\,Myr).  This last and
most significant error gives 0.075$\pm$0.006\,M$_\odot$ and
2914$\pm$43\,K for an object at the hydrogen burning limit and
1.000$\pm$0.027\,$M_\odot$ and 5270$\pm$70\,K for a solar type object.

\subsection{\label{spec} Spectroscopic Data Reduction and Calibration}

\begin{figure}
  \epsscale{1.0} \plotone{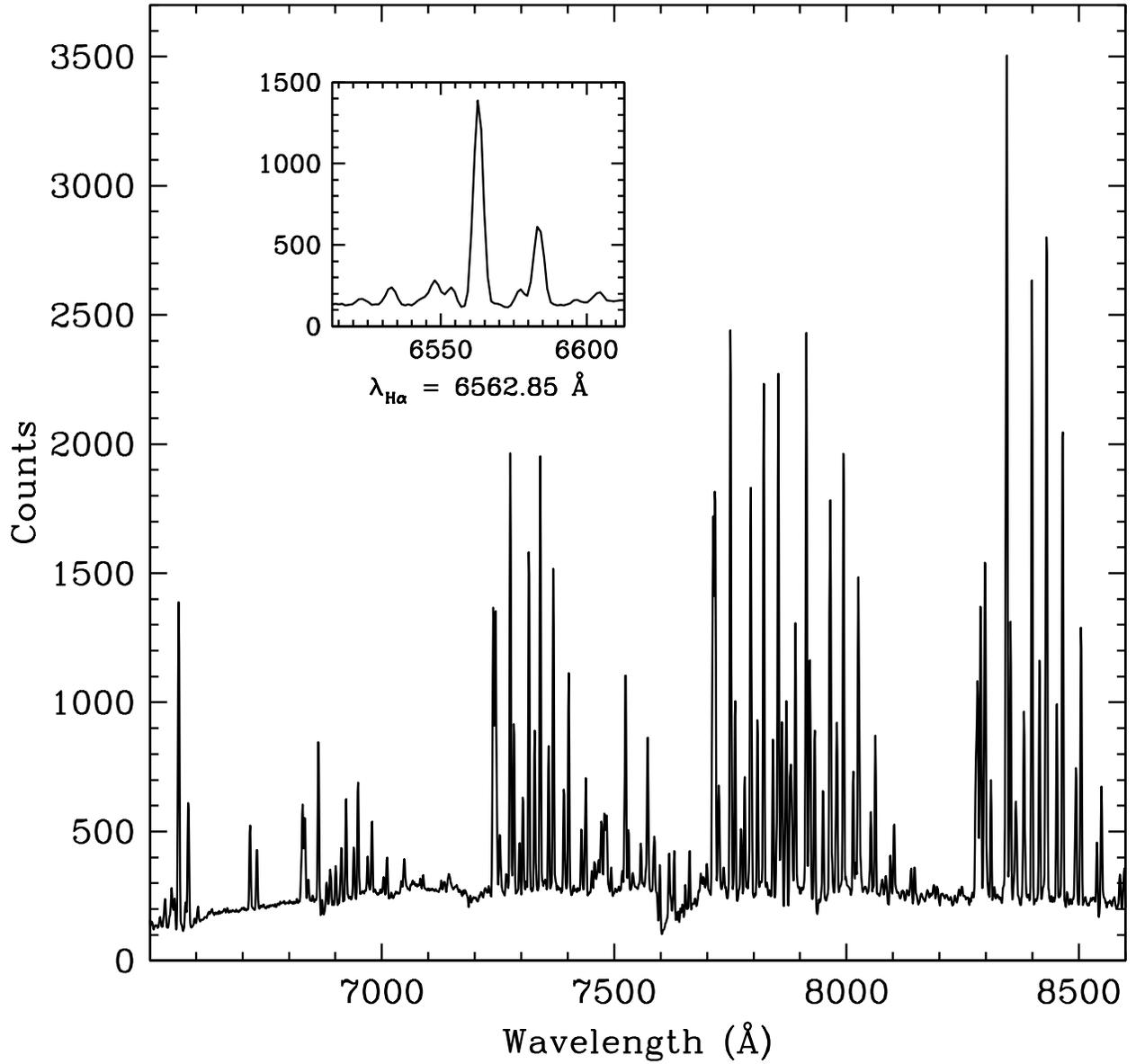}
  \caption{\label{fig:sky} Spectrum used for sky subtraction of our
    spectroscopic data. Note the H$\alpha$ (nebula) emission line with
    equivalent width of $W$(H$\alpha$)\,=\,48\,\AA.}
\end{figure}

The standard CCD reductions (overscan subtraction and trimming) were
performed on each image using the $ccdred$ package under IRAF.  We
then used the IRAF package $dohydra$ to perform flat-fielding (using
dome flats), throughput correction (with the skyflats) and scattered
light corrections. The spectra were wavelength calibrated using the
PENRAY comparison lamp with 2\,sec exposure time. Sky subtraction was
performed in a similar manner as fringe subtraction in photometry: a
\textit{standard} sky spectrum (shown at Figure \ref{fig:sky}) was
obtained from the median of our sky spectra (more than 20 fibers were
assigned for sky subtraction in each Hydra pointing) and scaled to
optimize the sky subtraction for each science spectrum individually.
However, this sometimes resulted in H$\alpha$ apparently being in
absorption for some objects.  We attribute this to H$\alpha$ emission
from the background itself. This is spatially variable and so
subtracting the sky spectrum (which includes H$\alpha$) sometimes
results in a over-subtraction of this feature. We discuss this
contamination problem and the danger of determining membership status
based on H$\alpha$ in \S \ref{h-alpha}.

Finally, flux calibration was performed with the spectrophotometric
standard Hiltner~600, which was observed three times a night, at three
different airmass.

\subsection{\label{spt} Spectral type, effective temperature and mass
  determination}

For the objects for which spectra is available, we estimated in
addition the spectral type using the PC3 index from \cite{martin1999}.
The distinction between M--dwarf and background M-giants and
M-supergiants was achieved using a CaH index (\citealt{jones1973}). We
visually inspected all spectra in order to confirm the spectral type
and luminosity class estimation. We estimated a spectroscopic $T_{\rm
  eff}$ from the spectral type using the temperature scale of
\cite{luhman1999b} for objects between M1V to M9V and of
\cite{martin1999} for objects from L0V and later. We then use our
isochrone of IC~2391 to obtain the mass based on $T_{\rm eff}$.

\section{\label{selection} CANDIDATE SELECTION PROCEDURE}

The selection procedure discuss here concerns only our photometric
data while the discussion of the selection of our spectroscopic
candidates is done in \S \ref{spec-follow-up}. The candidate selection
procedure is as follows (and explained in more detail in the remainder
of this section).  Candidates were first selected based on
colour-magnitude diagrams (CMDs). A second selection was performed
using colour-colour diagrams.  Third, astrometry was used to remove
objects with high proper motion.  Finally, non-candidates were
rejected based on a discrepancy between the observed magnitude in
815/20 and the magnitude in this band computed with the isochrone of
IC~2391 and our estimation of $T_{\rm eff}$. To be a cluster member in
this work an object has to satisfy all four of these steps.

\subsection{\label{selection-1st} First Candidate Selection: CMDs }

\begin{figure}
  \epsscale{1.0} \plotone{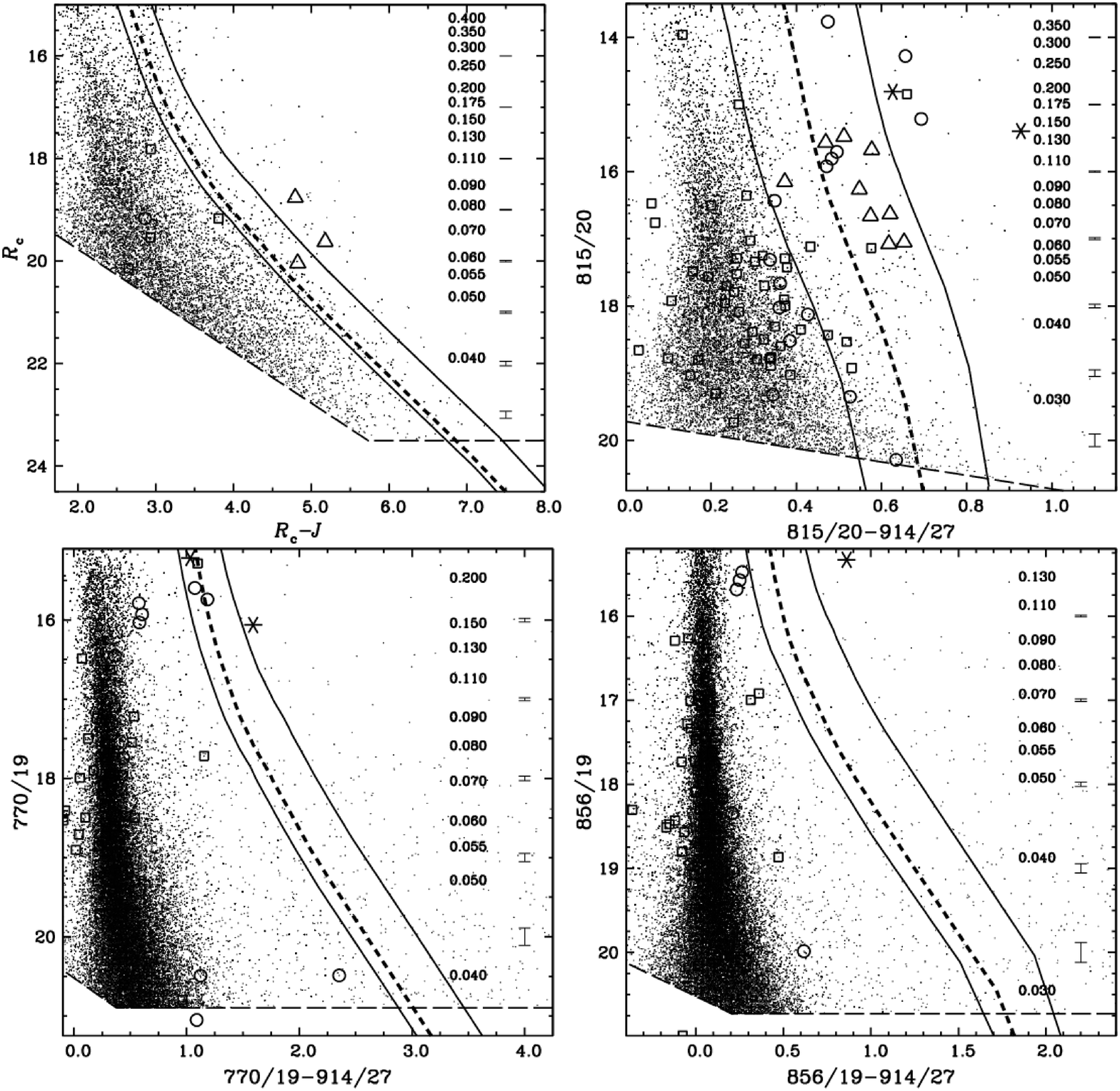}
  \caption{\label{fig:cmd} \textit{Top.} Two CMDs from the radial
    field 01. As dotted lines we show the isochrone computed from an
    evolutionary model with a grainless atmosphere (NextGen model, the
    masses for each $R_{\rm c}$ are shown in the right panel).  The
    thin dashed line is the 10$\sigma$ detection limit.  We also show
    candidate low mass cluster members from \cite{patten1999}
    (\textit{stars}), \cite{barrado2004} (\textit{triangles}),
    \cite{dodd2004} (\textit{squares}) and X-ray sources detected by
    XMM-Newton (\textit{circles}), which we detected in our survey.
    Some of these objects are not present in the left panel since the
    deep fields, where most of these objects are detected, lack
    $J$-band photometry. \textit{Bottom.} Two CMDs from the deep field
    32 with medium bands. Isochrones, 10$\sigma$ detection limit and
    cluster members from previous studies are the same as for the top
    two panels. In each panel, the thin solid lines represent the
    selection curve and the errorbars present the photometric errors.}
\end{figure}

Candidates were first selected from our CMDs by keeping all objects
which are no more than 0.15\,mags redder or bluer than the isochrone
in all CMDs (this number accommodates errors in the magnitudes and
uncertainties in the model isochrone), plus errors from age estimation
and distance to IC~2391 reflected on the isochrone. We additionally
include objects brighter than 0.753\,mag the isochrone in order to
include unresolved binaries. In Figure \ref{fig:cmd} we show two CMDs
for field 01 where candidates were selected based on 815/20 vs.
815/20--914/27 and $R_{\rm c}$ vs. $R_{\rm c}$--$J$ (\textit{top 2
  panels}).  We also present two CMDs from the deep field 32 using the
medium band 770/19, 856/14 and 914/27 (\textit{lower 2 panels}).  From
a total of 20~008~114 objects detected, 174~511 are kept (99.2\% are
rejected).

We also present in this figure low mass cluster member candidates from
previous work which we detected in our survey (\citealt{patten1999},
\citealt{barrado2004}, \citealt{dodd2004} and X-ray sources detected
by XMM-Newton where some are also presented in \citealt{marino2005}).
Some candidates from previous studies are simply not detected in our
work. This is the case with \cite{platais2007} where the faintest
candidates have $V$$\sim$15, which corresponds to 0.6\,M$_\odot$
(close to the saturation limit of the radial and outward fields at
$\sim$0.9\,M$_\odot$ and at the saturation limit of the central deep
fields, also at $\sim$0.6\,M$_\odot$). Also, no objects in our sample
match the 34 members studied by \cite{siegler2007} because either our
images are too deep so bright stars saturate (e.g.  HD74275, HD74374
and VXR22a, which saturate in the short exposures), or the objects are
not in our fields (e.g. VXR06, SHJM10 and PP07, which are between
fields 32 and 37).  Also, no objects match within 4 arcsec between our
objects and the 17 cluster candidates from \cite{rolleston1997} for
similar reasons: the bright stars saturate (e.g. object ID 162, 311
and 362, which are birghter then our saturation limit in $R_{\rm c}$),
or the objects are not in our fields (e.g. object ID 729 and 955,
which are in the central part of the cluster between fields 20 and
27).

Finally, we point out that the survey of IC~2391 based on proper
motion done by \cite{dodd2004} only covers an area of 1$^\circ$
diameter in the central part of the cluster.

\subsection{\label{selection-2nd} Second Candidate Selection:
  Colour-Colour Diagrams}

\begin{figure}
  \epsscale{1.0}
  \plottwo{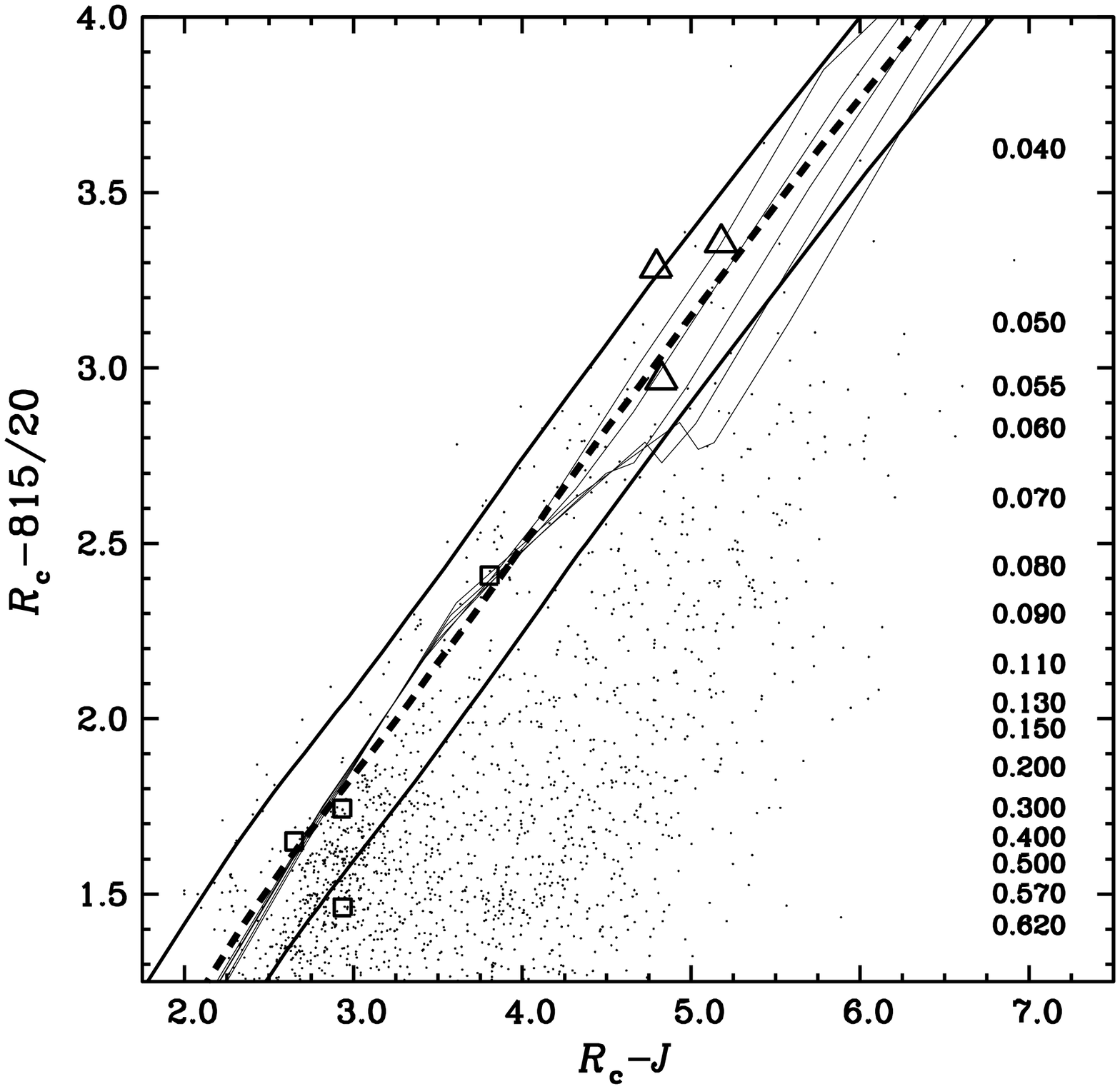}{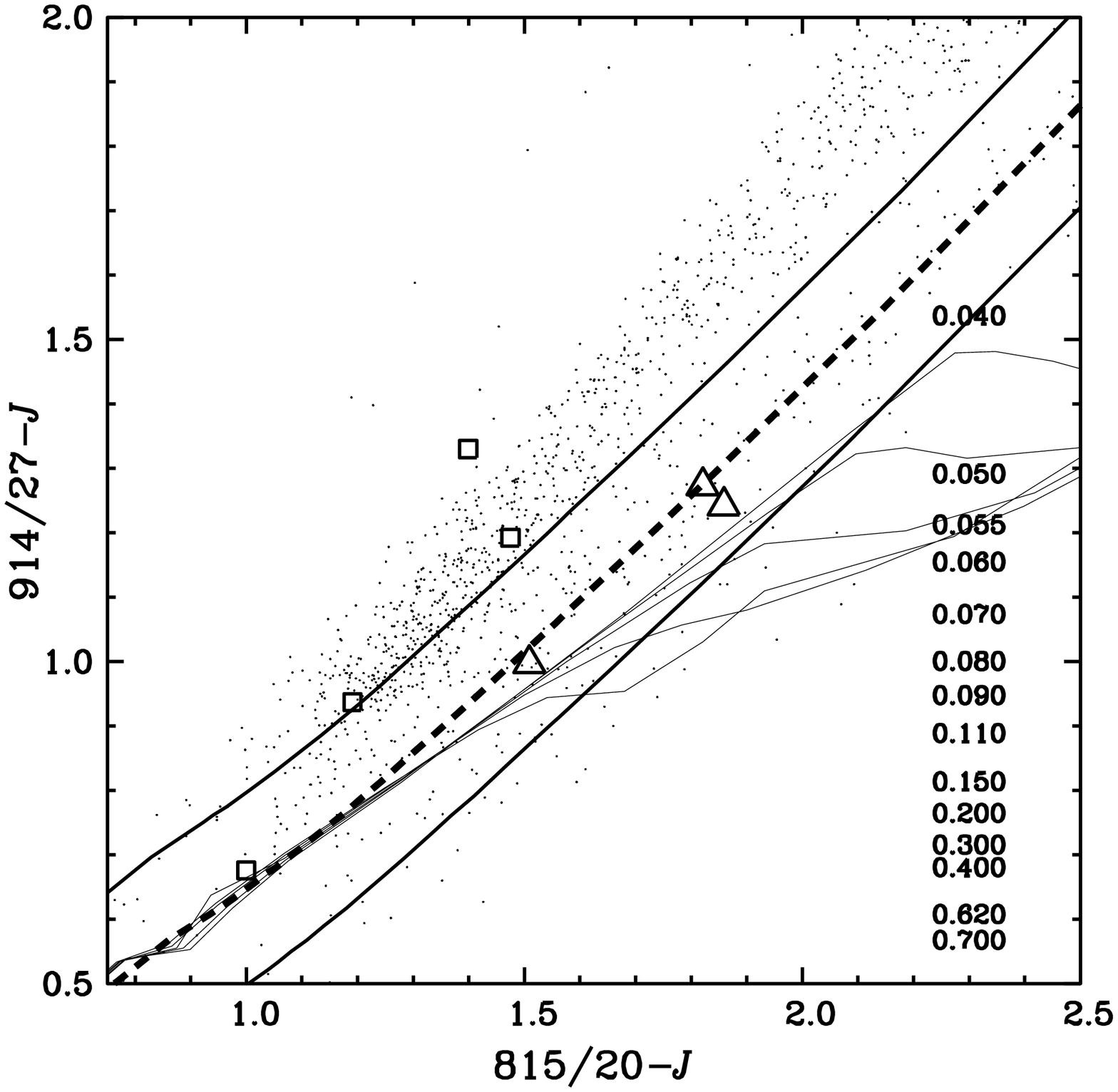}
  \caption{\label{fig:ccd} Two colour-colour diagrams of objects that
    are candidates based on our first selection (from field 01).
    Isochrones and objects from \cite{barrado2004}
    (\textit{triangles}), \cite{dodd2004} (\textit{squares}) and
    XMM-Newton (\textit{circle}) are as shown in Figure \ref{fig:cmd}.
    The thin lines represent the colour of possible background red
    giant contaminants. The colour-colour diagram on the left is
    therefore not suited for further candidate selection since the
    isochrone is spanned by the colour grid of the red giants (which
    is not the case for the colour-colour diagram on the right). In
    each panel, the thin solid lines next to the isochrone represent
    the selection curve.}
\end{figure}

\begin{figure}
  \epsscale{1.0} \plotone{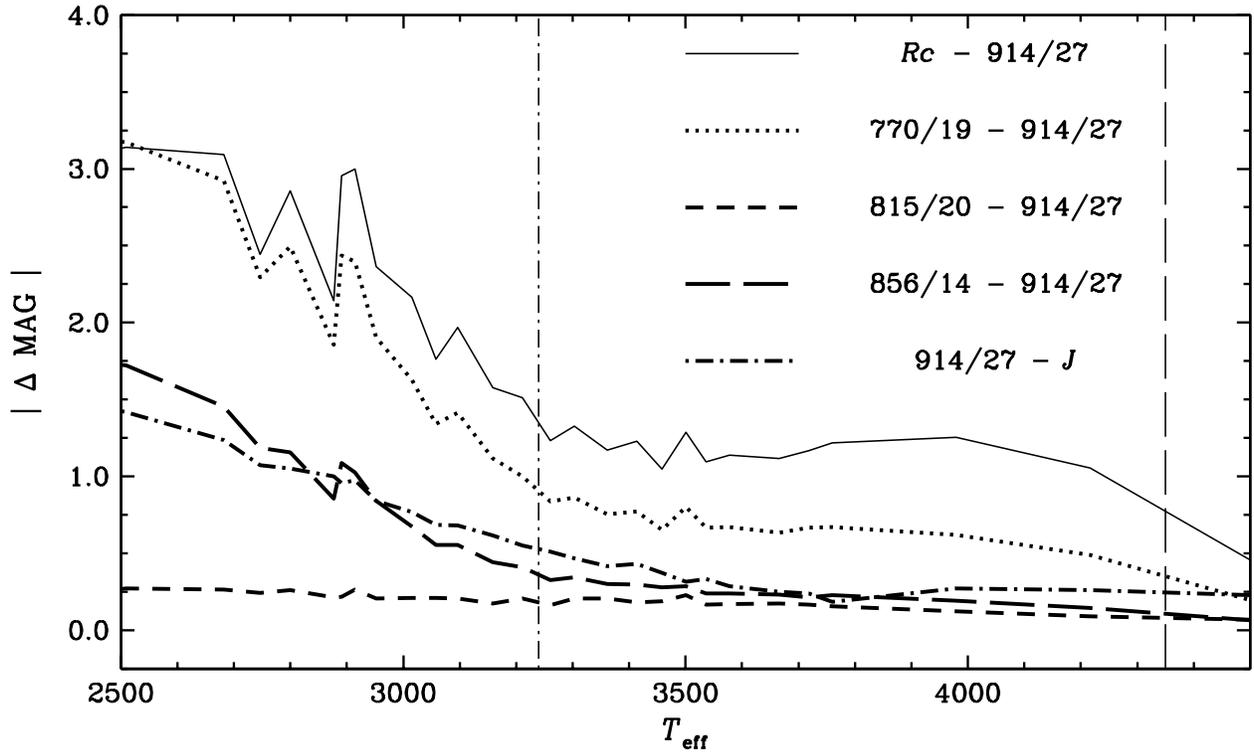}
  \caption{\label{fig:colour} Colour sensitivity to effective
    temperature for the medium band filter 914/27.  The vertical lines
    represents are at the approximate effective temperatures for
    spectral classes M5V (dash-dotted line) and K5V (long-dash line).
    We can see that there is no variation at all for 815/20-914/27
    compared to other colours at the L0V and M5V regime.}
\end{figure}

The second stage of candidate selection was achieved by taking all
objects within 0.15 mag of the isochrone of the NextGen model in
selected colour-colour diagrams. In Figure \ref{fig:ccd}, we present
two colour-colour diagrams where only the objects from the first
selection are plotted. Considering that many colour-colour diagrams
are possible (we have 4-5 filters), the variation of colour as a
function of $T_{\rm eff}$ was used to reject the use of colours for
which the NextGen model shows small variation in the M and L dwarf
regime (this is illustrated in Figure \ref{fig:colour} with the
815/20-914/27 colours).

Because one source of contamination are background red giants, we show
theoretical colours for such objects using the atmosphere models of
\cite{hauschildt1999b}, assuming that all objects have a mass of
5\,M$_{\odot}$, 0.5 $<$ log \textit{g} $<$ 2.5 and 2000\,K $<$ $T_{\rm
  eff}$ $<$ 6000\,K. We can see that $R_{\rm c}$--$J$ vs.\ $R_{\rm
  c}$--815/20 is not best suited for selecting candidates since the
isochrone is overlapped by red giant contaminants. However, in
815/20--$J$ vs.\ 914/27--$J$, we see a clear distinction between the
isochrone and the red giant contaminant in the brown dwarf regime (by
more than 0.2 mag). This procedure definitely helps to remove red
giant contaminants, and is further discussed in subsection
\ref{discussion-follow-up}. From a total of 174~511 objects, 33~794
are kept (80.6\% are rejected).

\subsection{\label{selection-3rd} Rejection of Contaminants Based on
  Proper Motion}

Although the RMS error of our astrometry is 0.15-0.20 arcsec (WFI) and
0.3-0.4 arcsec (CPAPIR), we nonetheless estimated proper motions per
year using the motion between the 1999/2000 WFI data and the 2007
CPAPIR data in an attempt to reject objects which deviate
significantly from the mean cluster proper motion in the literature.
The typical error on our proper motion measurement is 24 milliarcsec
per year (mas yr$^{-1}$).  The values of
($\mu_{\alpha}$cos$\delta$,$\mu_\delta$) for IC~2391 in mas yr$^{-1}$
from the literature are (-25.04$\pm$1.53,+23.19$\pm$1.23),
(-25.05$\pm$0.34,+22.65$\pm$0.28), (-24.64$\pm$1.13,+23.25$\pm$1.23)
and (-25.06$\pm$0.25,+22.73$\pm$0.22) from \cite{dodd2004},
\cite{loktin2003}, \cite{sanner2001} and \cite{robichon1999}
respectively. For our selection procedure we use the average of these,
(-25.0$\pm$2.0,+23.0$\pm$1.7).

\begin{figure}
  \epsscale{1.0}
  \plottwo{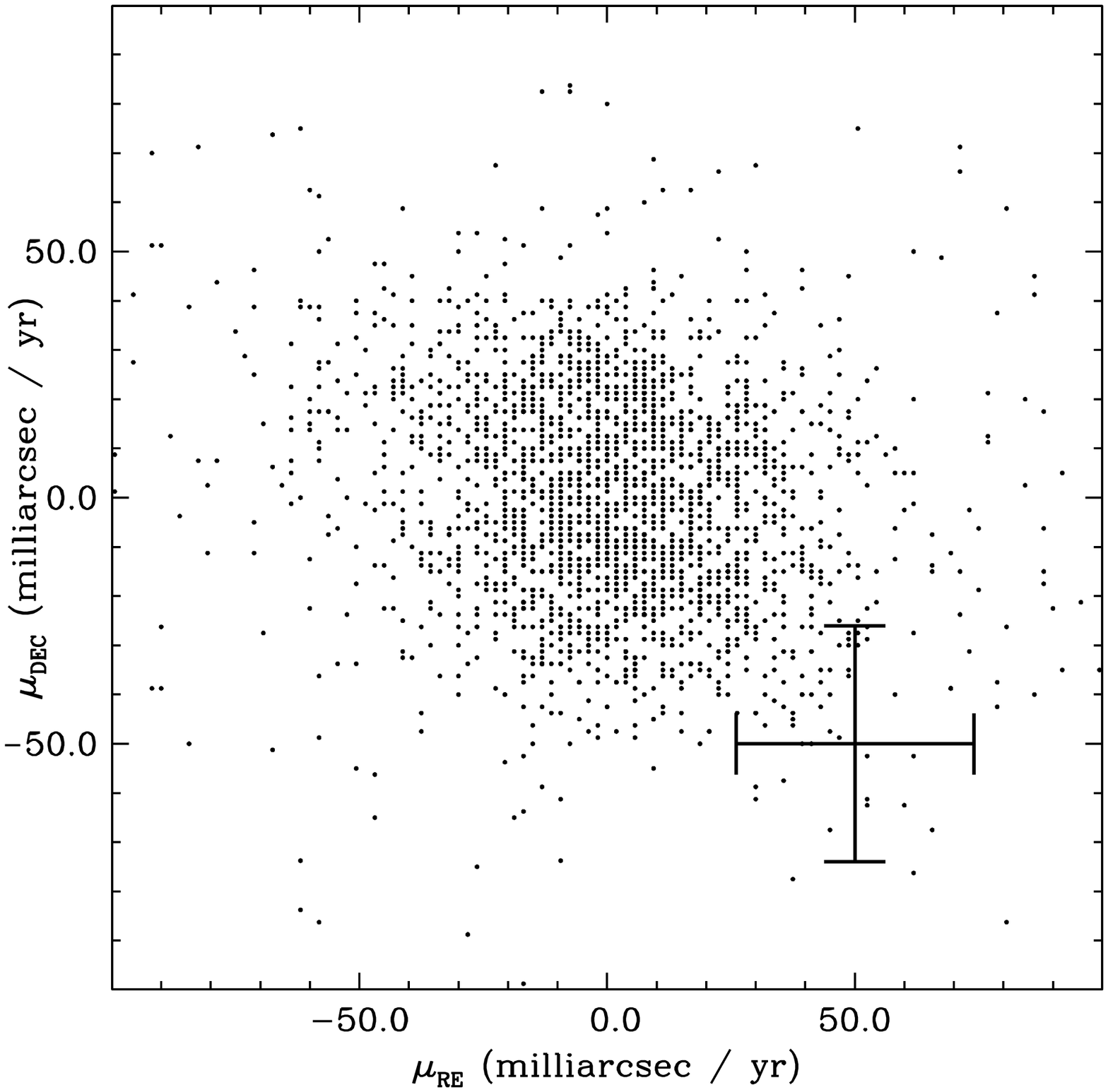}{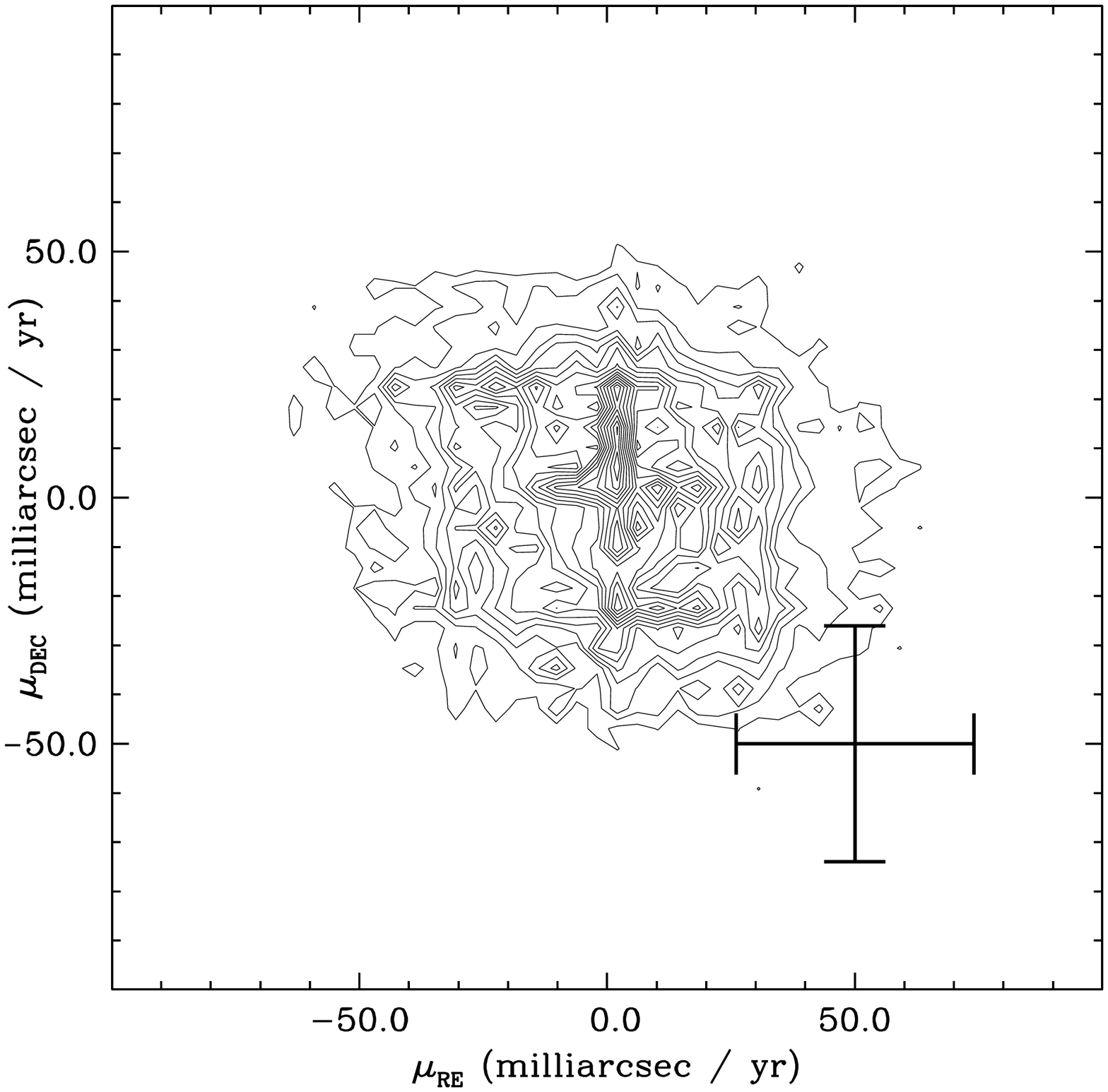}
  \caption{\label{fig:astrometry} \textit{Left.}  Proper-motion
    diagram from our survey (in milliarcsec per year). IC~2391 is at
    (-25.0,+23.0), which is taken from the literature (\S
    \ref{selection-3rd}). The typicaly error bar of individual objects
    in our survey is shown. For clarity, only one object out of five
    is shown. \textit{Right.}  Contour plot of the same data.}
\end{figure}

We first investigated whether the cluster itself could be identified
in the proper motion plane. To do this, we retained only those objects
detected from our observation runs with WFI (1999, 2000 and 2007) and
CPAPIR (2007) which have a match within 1\,arcsec.  We then examined
the distribution in the ($\mu_{\alpha}$cos$\delta$:$\mu_\delta$) plane
for any feature at (-25.0:+23.0). However, we see no clump in the
distribution of the proper motion (Figure \ref{fig:astrometry}).
Considering the large errors and the absence of any structure at the
expected location, we decided not to perform any selection using the
proper motion of IC~2391. However, astrometry is used to remove all
objects with a proper motion higher than 72 mas yr$^{-1}$ (3$\sigma$)
away from the cluster proper motion.

\subsection{\label{selection-4th} Rejection of Objects Based on
  Observed Magnitude vs.\ Predicted Magnitude Discrepancy}

\begin{figure}
  \epsscale{1.0} \plotone{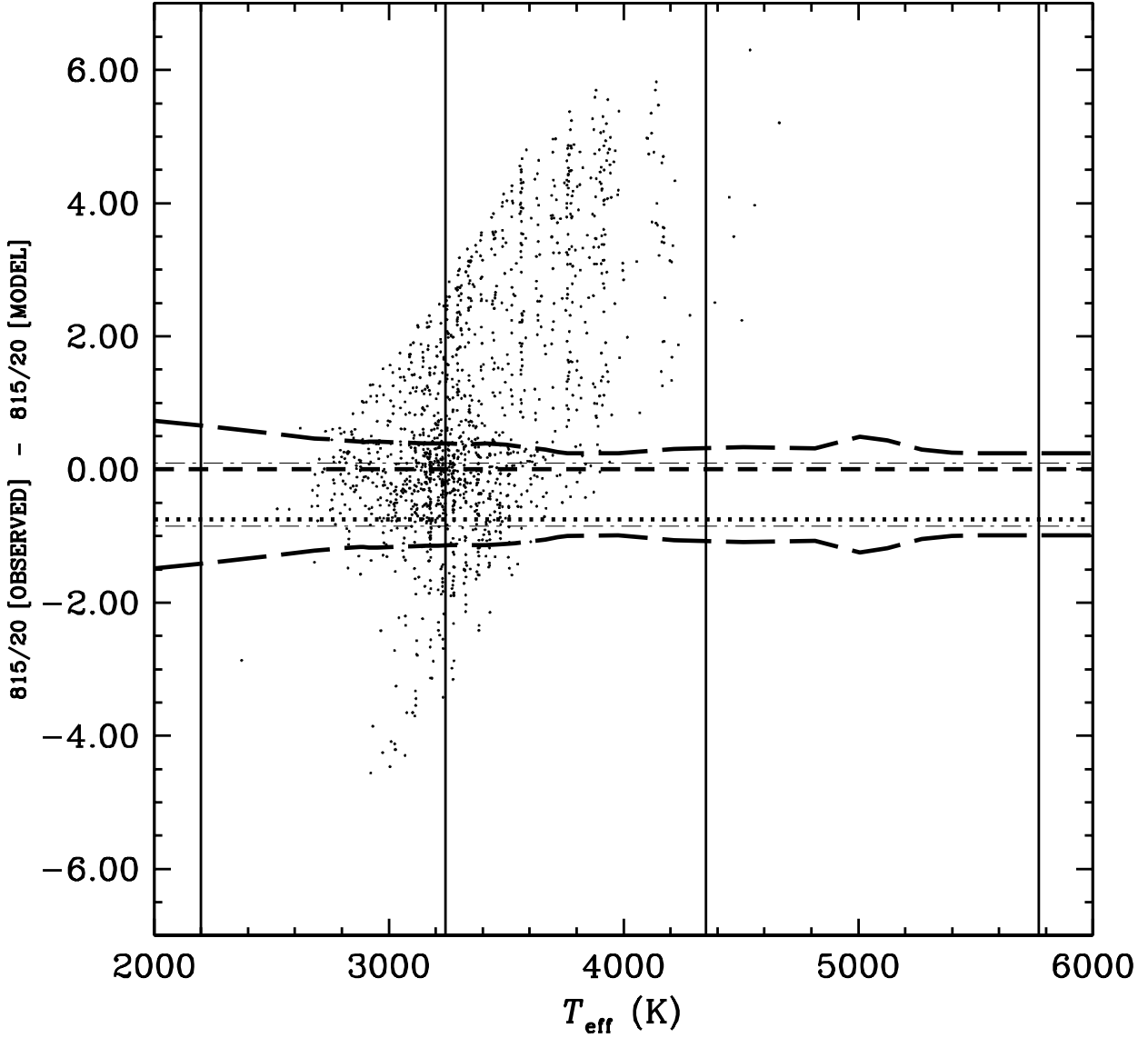}
  \caption{\label{fig:m816_vs_m816} Difference between the observed
    815/20 magnitude and that computed from the derived mass and
    $T_{\rm eff}$, as a function of $T_{\rm eff}$. The four vertical
    lines are at the positions of L0, M5, K5 and G5 dwarfs (left to
    right). The dotted line (at $-0.753$) represents the error due to
    the possible presence of unresolved binaries at, the dashed-dotted
    lines represents the error on the magnitude determination and the
    long dashed lines represents the uncertainties on the age and
    distance of IC~2391. (The short-dashed line just traces zero).}
\end{figure}

As indicated in \S \ref{get-mass}, our determination of $T_{\rm eff}$
is based on the energy distribution of each object and is independent
of distance. The membership status is determined by comparing the
observed magnitude of a given object in a band with the magnitude
predicted based on its derived $T_{\rm eff}$ and IC~2391's isochrone.
(The premise is that the predicted magnitude of a background
contaminant would be lower - brigher - than its observed magnitude and
higher for a foreground contaminant.) In order to avoid removing
unresolved binaries that are real members of the cluster, we keep all
objects with a computed magnitude of up to 0.753\,mag brigher than the
observed magnitude. In this procedure, we are also taking into acount
photometric errors and uncertainties in the age and distance
determinations of IC~2391. This is represented in Figure
\ref{fig:m816_vs_m816}. Combined with the rejection of contaminants
based on proper motion, this selection step reject 89.2\% of the
33~794 candidates obtained from the CMD and colour-colour diagrams.

\section{\label{results-survey} RESULTS OF THE SURVEY}

The final selection gives us 954 photometric candidates for outward
fields (namely fields 43, 46, 47, 48 and 49), 499 photometric
candidates for the four deep fields (15, 20, 27 and 32, with filters
$R_{\rm c}$, 770/19, 815/20, 856/14 and 914/27) and 1\,734 for all
other radial fields (observed with filters $R_{\rm c}$, 815/20, 914/27
and $J$). (We present in \S \ref{contamination-rate} a discussion of
the contamination of the radial fields.)  All our photometric
candidates are presented in Table \ref{tab:allphot}. Objects are given
the notation IC~2391-WFI-ZZ-YYY where ZZ is the field number and YYY a
serial identification number (ID).  Only the first 10 rows of the
tables are shown, the remainder available online. We also compare in
Table \ref{tab:allphot2} all objects in our sample which are also
confirmed as cluster members from \cite{barrado2004} and
\cite{dodd2004} and was detected by the X-ray Multi-Mirror Mission
(XMM-Newton). We see a good agreement between $T_{\rm eff}$ from our
photometric data and from \cite{barrado2004}, where only the colour
($R-I$)$_c$ was used to compute $T_{\rm eff}$.

Not all candidate members from previous studies, detected in our
survey, are members of IC~2391 based on our photometric selection. As
pointed above, a cluster member presented in this work is an object
that satisfies all four steps of our selection procedure. For
instance, among the two objects detected in our survey which are also
cluster candidates by \cite{patten1999}, one is recovered by our
selection (object number 8). From 10 objects classified as candidate
members from spectroscopy and photometry by \cite{barrado2004}, 5 are
recovered in our selection: objects CTIO-038 and -091 fail the
colour-colour diagrams test and objects CTIO-041, -049 and -091 fail
the predicted magnitude vs. observed magnitude test. One possible
source of disagreement could be the use of H$\alpha$ by
\cite{barrado2004} as a membership criteria (see \S \ref{h-alpha} for
further discussion of this issue).

A total of 53 objects classified as cluster members by \cite{dodd2004}
based on proper motion and photometry were detected in our survey.
However, one of Dodd objects is recovered in our survey (object number
155, also identified as CTIO-152). This is also the only matchs,
within 4 arcsec, with the cluster members of \cite{barrado2004}.
Considering the size of the window used for their proper motion
selection (in milliarcsec, -28$\geq$$\mu_{\alpha}$cos$\delta$$\geq$-20
and +20$\leq$$\mu_\delta$$\leq$+28) and the order of the error on the
known proper motion of IC~2391 ($\lesssim$2 miliarcsec in
$\mu_{\alpha}$cos$\delta$ and $\mu_\delta$, see \S \ref{selection-3rd}
below), one could suspect some contamination by field stars. This can
be confirmed by the large scatter in the CMDs of IC~2391 presented by
\cite{dodd2004} below $R$$\sim$12 and $K$$\sim$10 in Figure 3 and 4
respectively. Another survey on IC~2391 based on proper motion was
performed recently by \cite{platais2007}.  However, as discussed in \S
\ref{selection-1st}, no objects from this work were detected in our
survey.

Only two objects listed Table 2 from \cite{marino2005} (IC~2391
members observed with XMM-Newton) were also detected by our survey
(source number 86 and CTIO-130), but neither objects is recovered by
our photometric selection. The first one is a source that
overlaps\footnote{The EPIC cameras on XMM-Newton have an angular
  resolution of 6 arcsec. Two of the cameras are MOS (Metal Oxide
  Semi-conductor) CCD arrays (referred to as the MOS cameras) and on
  camera at the focus of this telescope uses pn CCDs (referred to as
  the pn camera).}  with VXR53 from \cite{patten1996} and was
identified as a suspected cluster member, and also overlaps with
CTIO-126 from \cite{barrado2001} and was classified as a cluster
member (however, there was no spectroscopic follow-up of CTIO-126 by
\citealt{barrado2004}).  This object is not recovered in our selection
because this candidate fail the observed magnitude vs.\ predicted
magnitude test. Another object presented by \cite{marino2005}
(observed only by the MOS cameras onboard XMM-Newton, but not by the
pn camera) is CTIO-130, but they noted that this star has $J$ and
($J$-$K$) values incompatible with the IC~2391 main sequence.

\subsection{\label{talk-ebv} Effect of Background Contamination on
  Candidate Selection}

\begin{figure}
  \includegraphics[scale=0.6,angle=-90]{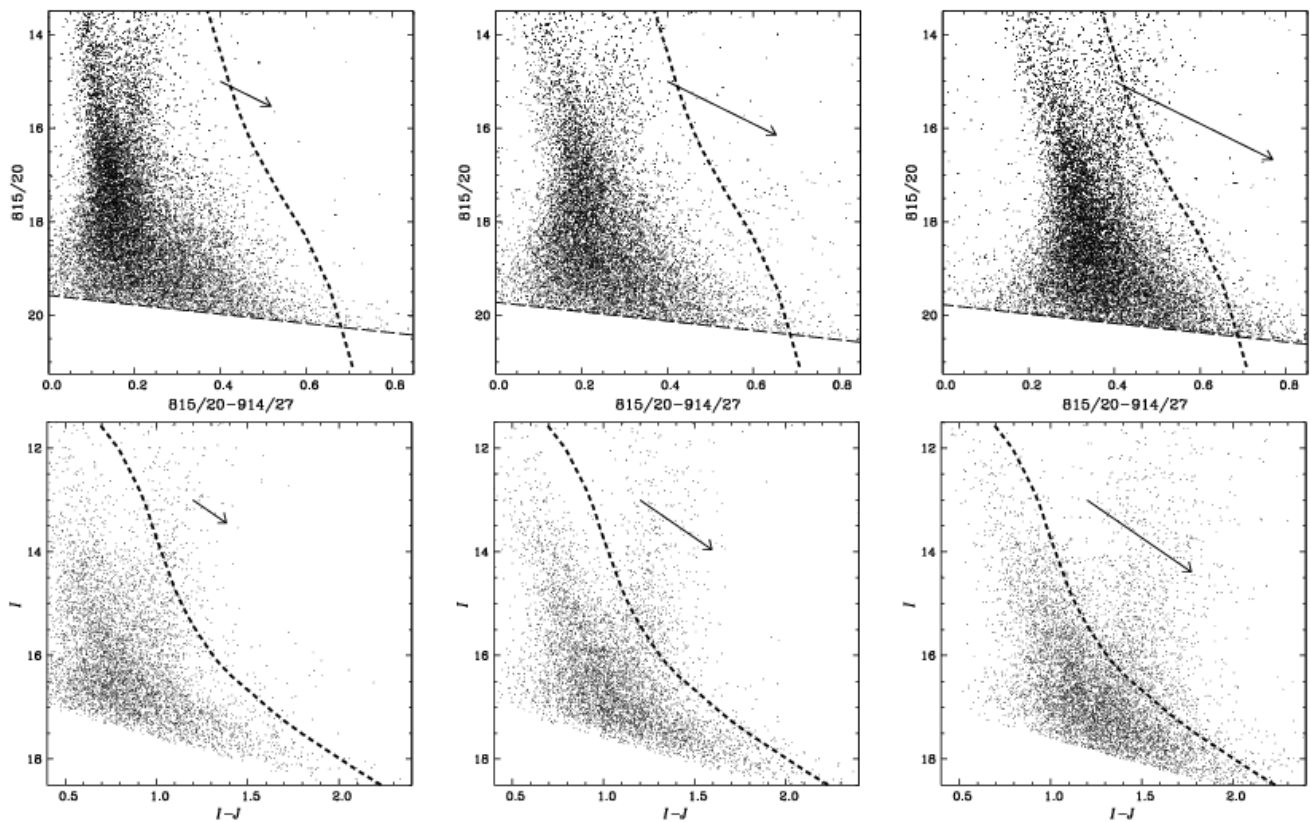}
  \caption{\label{fig:010940} CMDs of 3 fields from our survey
    (\textit{top}, from left to right, fields 40, 01 and 09) and 3
    CMDs from the same fields using DENIS data (\textit{bottom}, $I$
    versus $I$-$J$). The NextGen isochrones is also shown. We clearly
    see a colour shift of the (field star) locus between these fields
    in our data, as well in the DENIS data. For all panels, the arrows
    represents the reddening vectors based on $E$($B$--$V$) towards
    each fields.}
\end{figure}

In comparing the CMDs for different fields, we discovered something
peculiar (Figure \ref{fig:010940}, top 3 panels). We see a shift in
the colours of the bulk of the (field) stars from field to field,
something we also observe in other colours. The comparison of the
amplitude of this shift (for a given magnitude interval) with
observational parameters such as nights, airmass, seeing and
10$\sigma$ detection limit shows no correlation and there is no other
indication of reduction problems. We did, however, find a correlation
of the colour shift with the Galactic longitude \textit{b}.  However,
in order to verify that these shifts where real, we obtained DENIS
photometry (\textit{Deep Near Infrared Survey of the Southern Sky}) in
$I$ and $J$ band for the same fields presented in Figure
\ref{fig:010940}, which are field 01, 09 and 40. We can see that the
shift in the colours of the bulk of the (field) stars is also observed
in the DENIS data (Figure \ref{fig:010940}, lower 3 panels).

\begin{figure}
  \epsscale{1.0} \plottwo{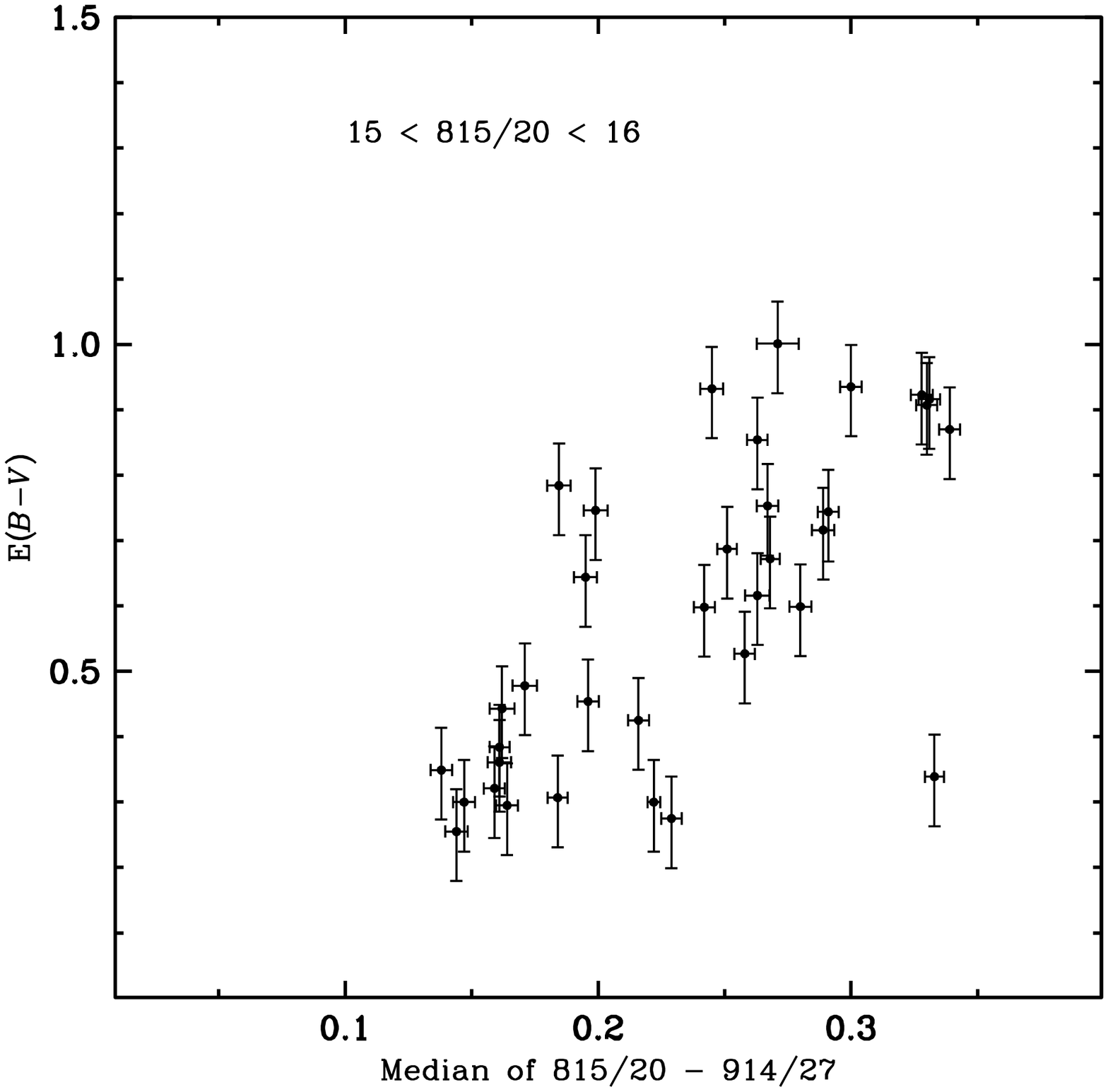}{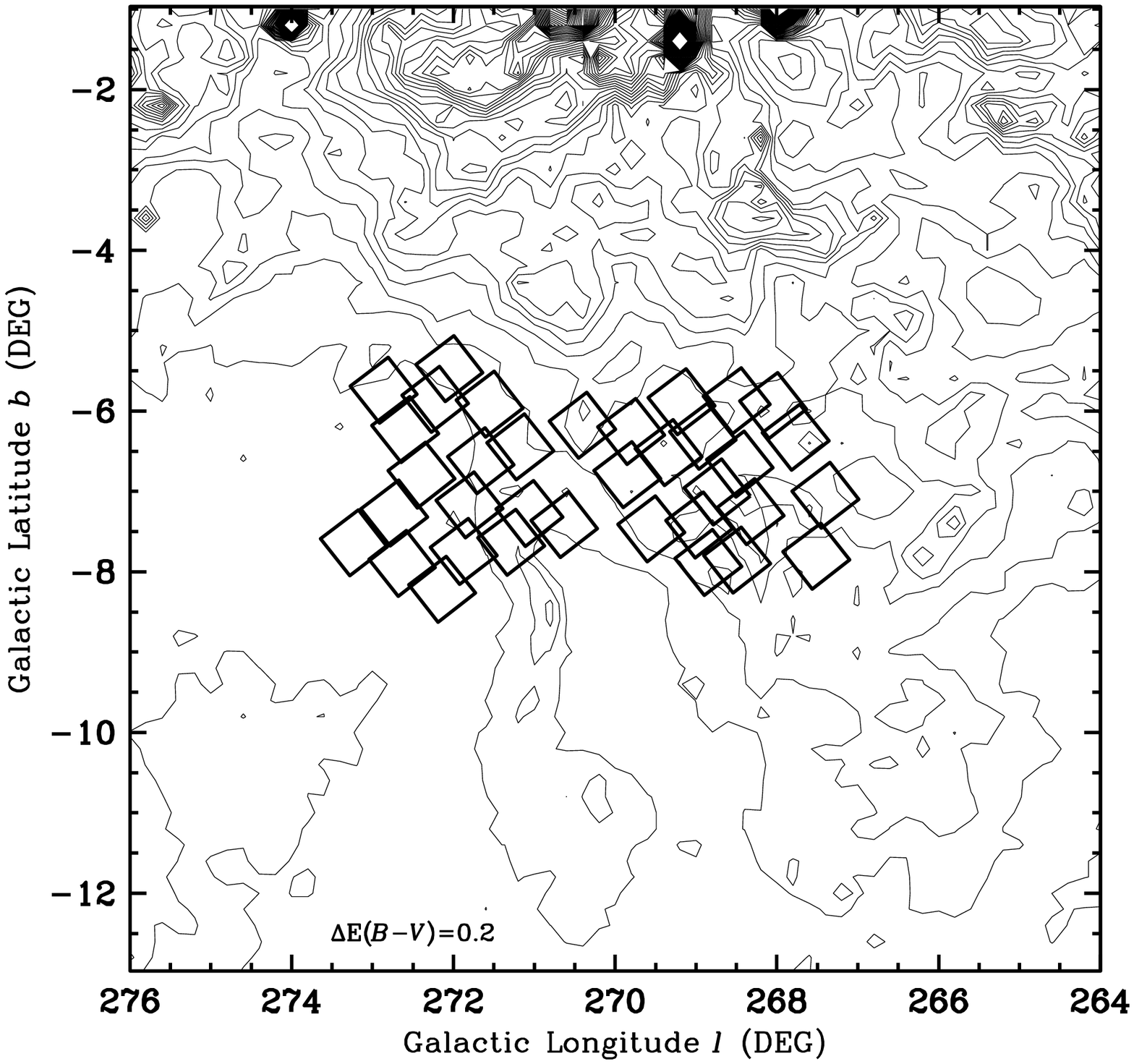}
  \caption{\label{fig:ebv} \textit{Left.} $E(B-V)$ towards all our
    fields (from the \citealt{schlegel1998} extinction map) plotted
    against the median 815/20-914/27 stellar colour in those fields.
    \textit{Right.} Position of the fields of our survey overplotted
    by the $E$($B$--$V$) extinction map of \cite{schlegel1998}. The
    contour separation is 0.2\,mag.}
\end{figure}

Although reddening is negligible for objects in IC~2391, this is not
the case for background objects, and these constitute most of the
stars in our sample. Due to the high variation of the background
extinction in this direction of the Galactic disk
(\citealt{schlegel1998}) -- the cluster is centered at
\textit{l}=270.4 \textit{b}=-6.9 -- some variation in the CMD locus
could be extinction-induced variations in the background stars. In
Figure \ref{fig:ebv} (\textit{left}), we plot the reddening $E$($B-V$)
in our fields against the median of the colour 815/20-914/27 (in a bin
of magnitude of 15 $<$ 815/20 $<$ 16) for all our fields. The colours
vary by as much as 0.25 mag. To better illustrate the spatial
variation of the background extinction, Figure \ref{fig:ebv}
(\textit{right}) shows the position of the fields of our survey
overplotted with the $E$($B$-$V$) extinction map of
\cite{schlegel1998}. This colour gradiant of the background stars has
not been reported in previous surveys of IC~2391 (\citealt{dodd2004};
\citealt{barrado2001}; \citealt{patten1999}). It can be expected that
\cite{barrado2001} and \cite{patten1999} didn't observed such shift in
colour since their survey cover a smaller area (2.5 and 0.8 sq.\ deg.
respectively) of the sky compared to our 10.9 sq.\ deg. coverage.

\subsection{\label{mf} Mass Function}

The mass function, $\xi$(log$_{10}$M), is generaly defined as the
number of stars per cubic parsec (pc$^3$) in the logarithmic mass
interval log$_{10}$M to log$_{10}$M + $d$log$_{10}$M. Here, we do not
compute the volume of IC~2391 so instead we present a MF using the
total number of objects in each 0.1 log$_{10}$M bin per 1~000
arcmin$^2$, starting at the mass bin log$_{10}$M=-1.65
($\sim$0.02\,M$_\odot$). The mass functions computed here are all
system mass functions since we don't make any corrections for
binaries. We analyse the radial variation of the MF using the fields
with photometry with the filters $R_{\rm c}$, 815/20, 914/27 and $J$
(Figure \ref{fig:mf-r89j}). Mass functions were computed over three
regions: for fields between 0.5$^\circ$ to 1.5$^\circ$ of the cluster
center (which corresponds to 1.3\,pc and 3.8\,pc respectively); for
the annulus from 1.5$^\circ$ to 2.1$^\circ$ (which corresponds to
5.4\,pc); for fields outside of 2.1$^\circ$\footnote{For reference,
  the core and tidal radius of IC~2391 estimated by
  \citealt{piskunov2007} are 1.2\,pc and 7.38\,pc respectively, which
  corresponds to 0.35$^\circ$ and 2.89$^\circ$ from the cluster
  center}.  We have also computed a MF for all fields within
2.1$^\circ$ of the cluster center to help radial variation analysis
and present this as our estimation of the MF for IC~2391.
Furthermore, we have measured the MF for the five outward fields and
also for the four deep fields (Figure \ref{fig:mf-deep-89j-fit}).

Since the radial fields were also observed with $R_{\rm c}$ in
addition to the filters used for both the radial and outward fields,
we use this to estimate the (additional) contamination in outward
fields relative to the radial fields for each mass bin.  To do so, we
performed a photometric selection for our radial fields using only the
filters available in the outward fields (815/20, 915/27 and $J$).  We
compared the MF computed from this with the MF from the outward fields
and obtained, for each mass bin, the number of object that would have
been rejected if we would had an additional $R_{\rm c}$-band
observation. (Here we make the assumption that the true MF should be
the same in the radial and outwards fields.)

In Figure \ref{fig:mf-deep-89j-fit} (left panel) we present the
\textit{uncorrected} MF of the outward fields and the
\textit{corrected} MF of the outward fields (right panel). It is not
possible to perform such corrections for the deep fields.

Useful (and simple) parametrizations of the mass function include the 
power law of \cite{salpeter1955} and a lognormal
\begin{equation}
  \xi(\textrm{log}_{10}\,\textrm{M})=k\cdot\textrm{exp}{\biggl[-\frac{(\textrm{log}_{10}\,\textrm{M}-\textrm{log}_{10}\,\textrm{M}_0)^2}{2\sigma^2}\biggr]},
\end{equation}
\noindent where $k$=0.086, $m_0$=0.22\,M$_\odot$ and $\sigma$=0.57 was
derived for the Galactic field by \citealt{chabrier2003}).  Fitting
the lognormal mass function to our data for all fields within
2.1$^\circ$ of the cluster center, we obtain $k$=10.7$\pm$3.2,
$m_0$=0.13$\pm$0.03\,M$_\odot$ and $\sigma$=0.46$\pm$0.07. This is
overplotted in Figure \ref{fig:mf-deep-89j-fit}.

If we assume that the lognormal fit of the MF describe the behaviour
of the population in the mass range 0.02--0.9\,M$_\odot$ in IC~2391,
the total number of object expected is 3~985 for a total mass of
679\,M$_\odot$. (In \S \ref{contamination-rate} we will discuss again
the total number of object and total mass, following an estimation of
the contamination for each mass bin in the MF of the radial fields.)

We present in Figure \ref{fig:mf-r89j2} the MF for all fields within
2.1$^\circ$ of the cluster center and from other open clusters with
similar ages (NGC~2547, $\sim$30\,Myr; IC~4665, 28$^{\rm +7.3}$$_{\rm
  --6.6}$)\,Myr based on Li depletion boundary, \citealt{manzi2008}).
We also show on Figure \ref{fig:mf-r89j2} the MF of IC~2391 as
determined in previous work (i.e. from \citealt{barrado2004} and
\citealt{dodd2004}). All were normalized to the Galactic field star MF
at 0.3\,M$_\odot$.

\section{\label{mf-variation} ANALYSIS AND DISCUSSION OF THE MASS
  FUNCTIONS}

In the following subsection, we discuss the mass function derived from
the deep fields and from outward fields only. The other fields are
used to study the radial variation of the MF and are subject of
further discussion in the following two subsections. We complete this
section with a discussion of the contamination rate by non-cluster
members.

\subsection{\label{mf-deep-43-46-47-48-49} Mass Function of the
  outward fields and of the deep fields}

Considering the fact that only three bands were used for the outward
fields, and thus fewer constraints imposed, we expect that the number
of photometric candidates would be larger per unit area than the other
fields.  The MF (Figure \ref{fig:mf-deep-89j-fit}, left panel) shows
more low mass objects (compared to the MF of the radial field, for
masses below $\sim$\,0.15\,M$_\odot$), a similarity with the radial MF
from 0.13 to 0.3\,M$_\odot$, and again more stellar objects in the
mass range 0.5 to 0.8\,M$_\odot$. The \textit{corrected} MF (Figure
\ref{fig:mf-deep-89j-fit}, right panel) shows a better agreement with
the radial MF.

The MF from the deep fields (Figure \ref{fig:mf-deep-89j-fit}) agrees
with the mass function of the radial fields within 2.1$^\circ$ from
cluster center in the mass range 0.05 to 0.1\,M$_\odot$ and above
0.2\,M$_\odot$.  However, there is more substellar objects below
0.05\,M$_\odot$.

The rise of the MF for objects below 0.05\,M$_\odot$ was also observed
in IC~2391 by \cite{barrado2004}. In their work, the MF was computed
with objects that were selected as cluster members based on $R_{\rm
  c}$, $I_{\rm c}$, $J$, $H$ and $K$ photometry. Since their NIR
photometry was taken from 2MASS, no data are available for objects
fainter than $I_{\rm c}$$\lesssim$19 (10$\sigma$ detection limit of
2MASS is at $J$$\sim$15.8). As a result, their selection for fainter
objects was based on $R_{\rm c}$ and $I_{\rm c}$ photometry only. In
our case, although $J$ band photometry is available for the outward
fields, no $R_{\rm c}$ photometry is available. Thus
\cite{barrado2004} used a relatively short baseline ($R_{\rm
  c}$--$I_{\rm c}$) for their selection, as did we in our outward
fields (815/20, 914/27 and $J$), both of which are considerably
shorter than the baseline we used in the radial fields ($R_{\rm c}$,
815/20, 914/27 and $J$).  This situation is also observed in the MF of
the deep fields (where $R_{\rm c}$ band photometry is available, but
no $J$ band photometry).  Only the fields observed with $R_{\rm c}$
and $J$ as well (i.e. a longer baseline) show no significant rise of
the MF below 0.05\,M$_\odot$ (Figure \ref{fig:mf-deep-89j-fit}, left
panel).  Since no red giants were observed in our spectroscopic
follow-up of the two deep fields 15 and 20 (\S \ref{spec-follow-up}),
we conclude that this increase is an artefact due to contamination by
M--dwarfs. In this low mass regime (for objects with mass
$\lesssim$0.05\,M$_\odot$), a long spectral baseline (including, for
instance, $R_{\rm c}$ and $J$) is needed to efficiently remove
contaminations, as it allows a better determination of the energy
distribution. This is confirmed when we compare the corrected and
uncorrected MFs of the outward fields (Figure
\ref{fig:mf-deep-89j-fit}).

The rise in the MF over 0.5--1.0\,M$_\odot$ is observed in the outward
fields but not in the deep fields (Figure \ref{fig:mf-deep-89j-fit}).
\cite{jeffries2004} present the MF of the open cluster NGC~2547 and
also noticed a rise in the 0.7--1.0\,M$_\odot$ interval (Figure
\ref{fig:mf-r89j2}). This rise, also observed in the luminosity
function as a large peak at 12\,$\lesssim$\,$I_{\rm
  c}$\,$\lesssim$\,14.5, they attribute to contaminating background
giants. This would be consistent in the fact that we see this in the
radial fields (see Figure \ref{fig:mf-r89j}) but not in the deep
fields (Figure \ref{fig:mf-deep-89j-fit}). Indeed, as we discuss later
in \S \ref{spec-follow-up}, no red giants were found in our
spectroscopic follow-up, confirming that the use of medium bands such
as 770/19, 815/20, 856/14 and 914/27, and theoretical colours
(\citealt{hauschildt1999b}) are effective in removing background red
giants. However, from the MF of the radial fields (including the
outwards fields), the medium filters 815/20 and 914/27 alone, combined
with wide band $R_{\rm c}$ and/or $J$, are less efficient at removing
background giants. Therefore, the rise in the MF of the outward fields
over 0.5--1.0\,M$_\odot$ is related to the filters used, but is not a
baseline issue.

We also observe on Figure \ref{fig:mf-deep-89j-fit} that two mass bins
at $\sim$0.11 and $\sim$0.18\,M$_\odot$ are significantly high
compared to the other mass bins in the mass range
0.05--0.3\,M$_\odot$.  Considering that this is not observed in the MF
of the radial field and in the corrected/uncorrected MF of the outward
fields, we suspect that this is due to the selection procedure related
to the deep fields so no conclusion should be made based on these two
mass bins.

\subsection{\label{mf-variation-vlms} Radial Variation of the Mass
  Function at the Stellar and at the Substellar Regimes}

\begin{figure}
  \epsscale{1.0} \plotone{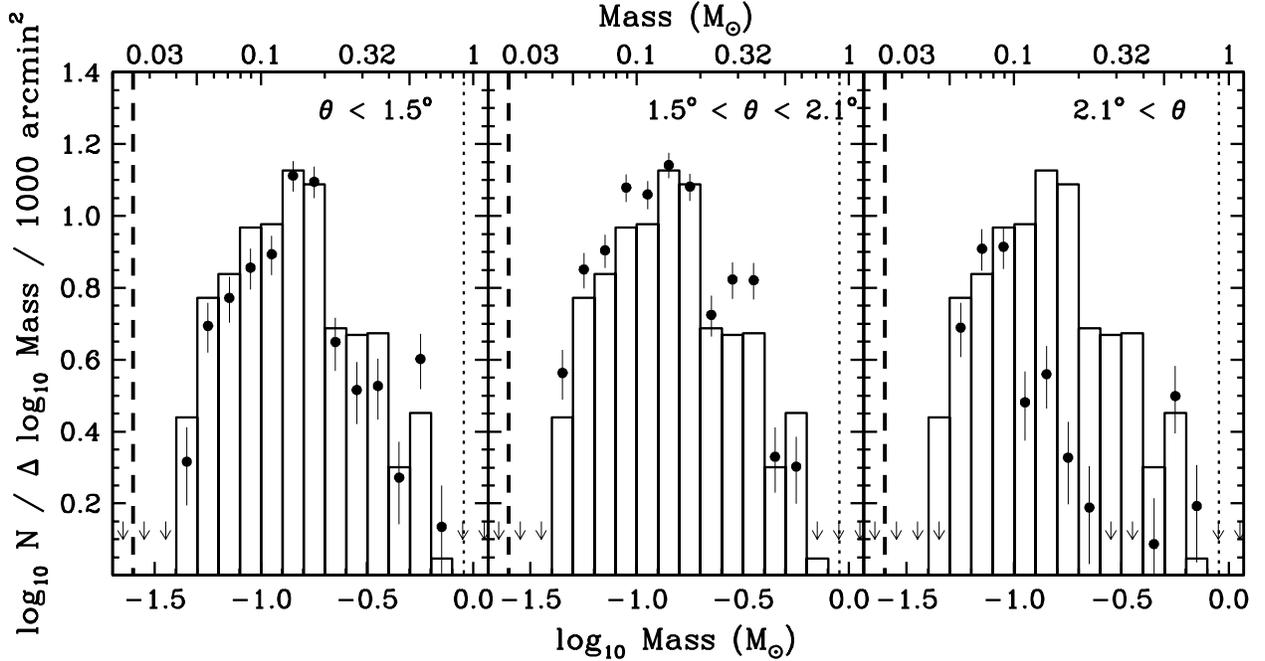}
  \caption{\label{fig:mf-r89j} MF based on photometry for all radial
    fields. The 10$\sigma$ detection limit is shown as a vertical
    dashed line. Dots in each panel represent the MF of
    (\textit{left}) fields within 1.5$^\circ$ of the cluster center,
    (\textit{center}) fields within the annulus from 1.5$^\circ$ to
    2.1$^\circ$ and (\textit{right}) the MF of fields outside of
    2.1$^\circ$.  Error bars are Poissonian arising from the number of
    objects observed in each bin. The histogram is the MF for all
    fields within 2.1$^\circ$ of the cluster center.  The vertical
    thin dotted line is the mass for which saturation start to occur
    in the short exposures. (The total area covered for each panel,
    from left to right, is 6~637, 9~539 and 5~609 arcmin$^2$.)  Just
    for reference, the ordinate value of 1.13 for the the bin at
    log$_{10}$M=-0.85 (0.14\,M$_\odot$) -- the histogram peak --
    corresponds to 109 objects.}
\end{figure}

\begin{figure}
  \epsscale{1.0} \plottwo{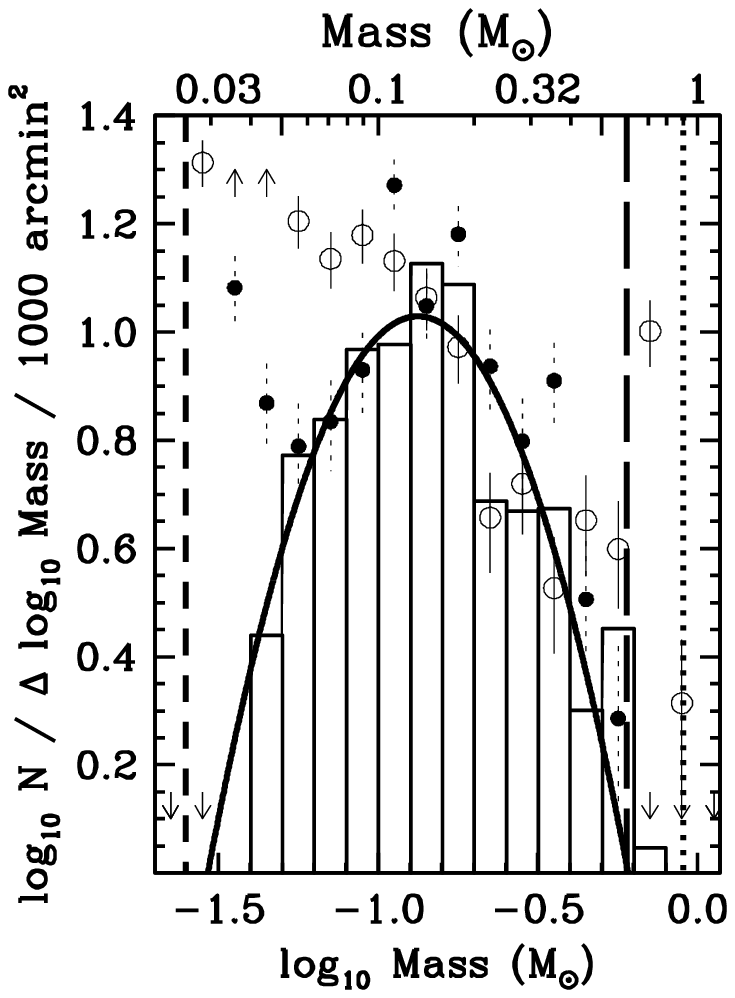}{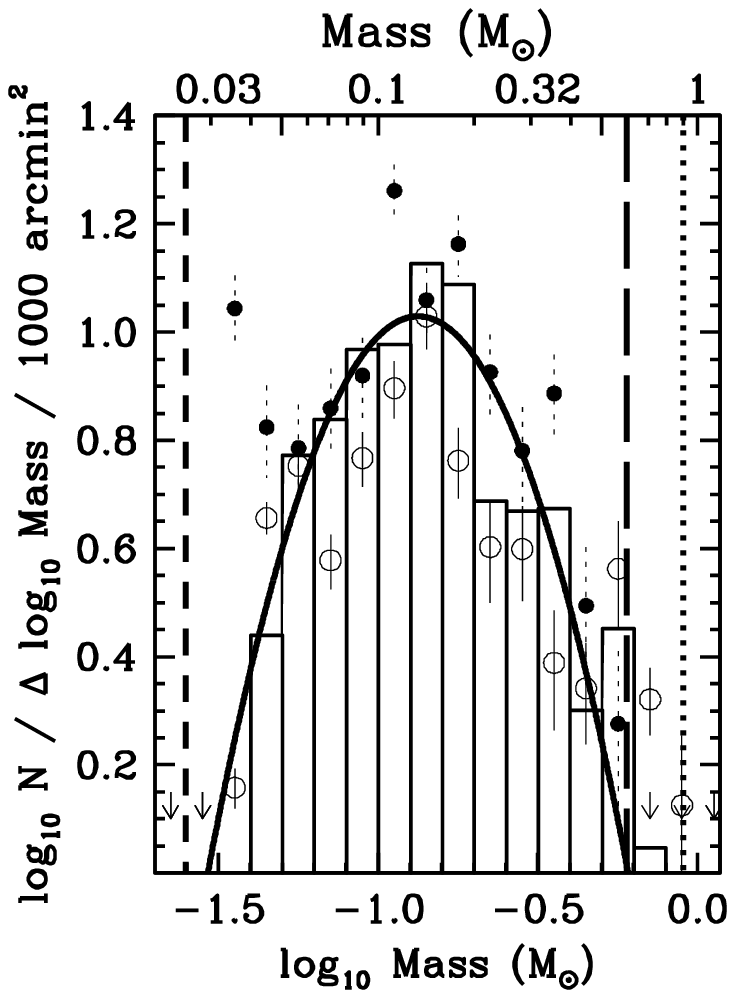}
  \caption{\label{fig:mf-deep-89j-fit} \textit{Left.} Filled dots
    represent the MF based on the four deep fields (observed with the
    wide bands $R_{\rm c}$ and the medium band 770/19, 815/20, 856/14
    and 914/27) and open dots represent the MF based on the outward
    fields (observed with the wide band $J$ and the medium bands
    815/20 and 914/27). Also, we present the 10$\sigma$ detection
    limit, the MF of all fields observed in $R_{\rm c}$, 815/20,
    914/27 and $J$ within 2.1$^\circ$ from the cluster center and its
    log normal fit.  The vertical thin dotted and thin dashed line
    lines are the mass for which saturation start to occur in the
    short exposures for outward and deep field respectively.
    \textit{Right.} Same as the left panel, but the outward fields MF
    presented is the \textit{corrected} MF.}
\end{figure}

\begin{figure}
  \epsscale{0.5} \plotone{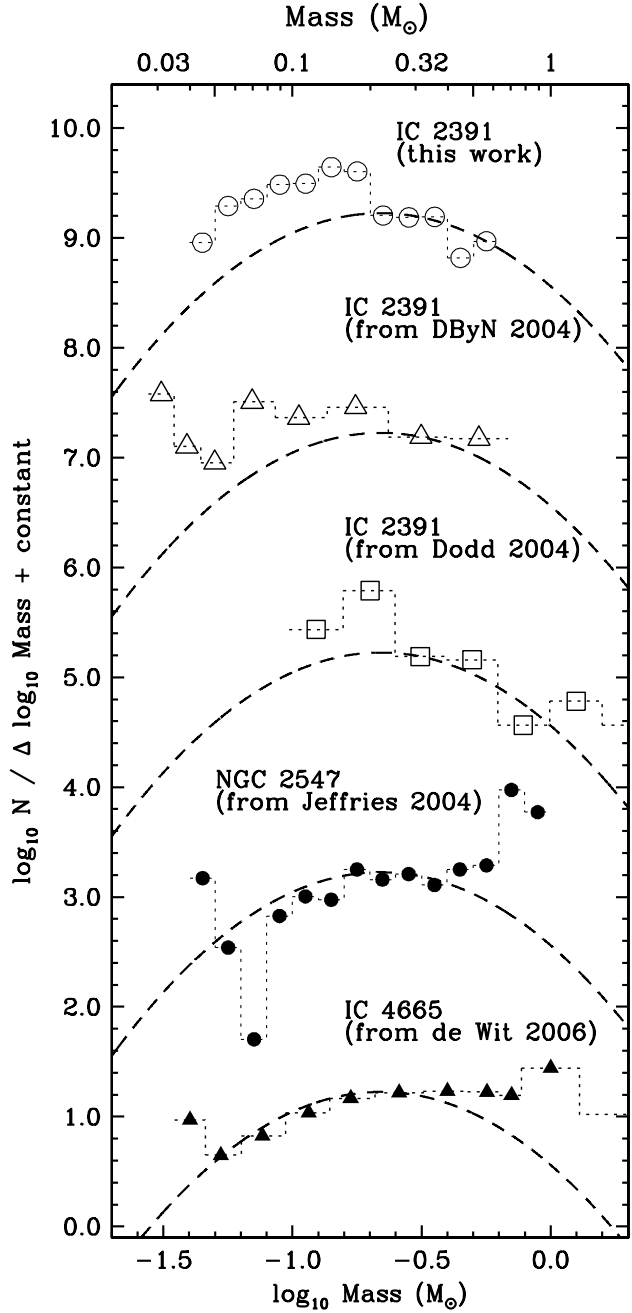}
  \caption{\label{fig:mf-r89j2} MF of IC~2391 from our present work
    (\textit{empty dots}, from all fields within 2.1$^\circ$ from the
    cluster center), for IC~4665 (\textit{filled triangles}) from
    \cite{dewit2006} and for NGC~2547 (\textit{filled dots}) from
    \cite{jeffries2004}.  We also show the galactic field stars MF
    from \cite{chabrier2003} as a dashed line. We also present the MF
    of IC~2391 from \cite{barrado2004} (\textit{empty triangles}) and
    from from \cite{dodd2004} (\textit{empty squares}).}
\end{figure}

From a first glance at Figure \ref{fig:mf-r89j}, we see that the two
mass functions within 2.1$^\circ$ (the points in the two left-hand
panels) are somewhat similar.  While there is some differences, these
are not very significant compared to the difference between their
common MF within 2.1$^\circ$ (plotted as the histogram) and the MF in
the outskirts of IC~2391 (beyond 2.1$^\circ$) as shown in the
right-hand panel.  Indeed, the MF for $\theta$$>$2.1$^\circ$ shows a
significant deficiency of stellar objects from 0.1 to 0.3\,M$_{\odot}$
(log$_{10}$M=-0.6) compared to the MF from the inner part of the
cluster, whereas outside of this mass range there is no significant
change with radius. Although we observe a number of objects in the
highest mass bins at 0.5--0.7\,M$_\odot$ (log$_{10}$M=-0.3 to -0.15),
we concluded (in \S \ref{mf-deep-43-46-47-48-49}) that this range of
masses is subject to significant contamination by red giants, so no
conclusion should be drawn from the radial variation of the MF in this
mass interval.

\begin{figure}
  \epsscale{0.5} \plotone{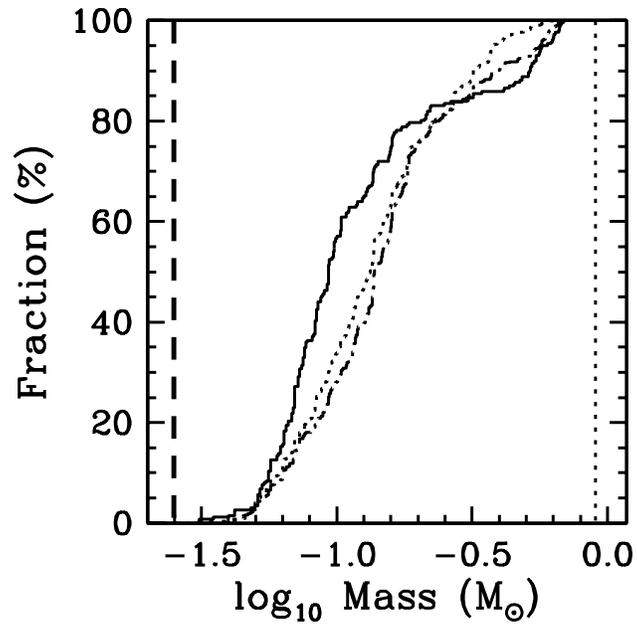}
  \caption{\label{fig:mass_func_89jr_cumulative} Cumulative number of
    cluster members within 1.5$^\circ$ (dash-dotted line), within the
    annulus from 1.5$^\circ$ to 2.1$^\circ$ (dotted line) and outside
    of 2.1$^\circ$ (tick line).  The 10$\sigma$ detection limit is
    shown as an horizontal dash line.  The vertical thin dotted line
    is the mass for which saturation start to occur.}
\end{figure}

In Figure \ref{fig:mass_func_89jr_cumulative} we present the
cumulative mass function for the same three regions of the cluster
presented in Figure \ref{fig:mf-r89j}. We again see the relative
absence of objects from 0.1 to 0.3\,M$_{\odot}$ (log$_{10}$M=-0.6) for
the inner radii.  A Kolmogorov-Smirnov test performed on these
distributions indicates that there is only a 1.1$\cdot$10$^{-5}$\,\%
probability of getting such a difference under the null hypothesis
that the population at $\theta$$>$2.1$^\circ$ is the same as that at
$\theta$$<$2.1$^\circ$, thus reinforcing the suggestion that the mass
functions are significantly different.

\begin{figure}
  \epsscale{1.0} \plotone{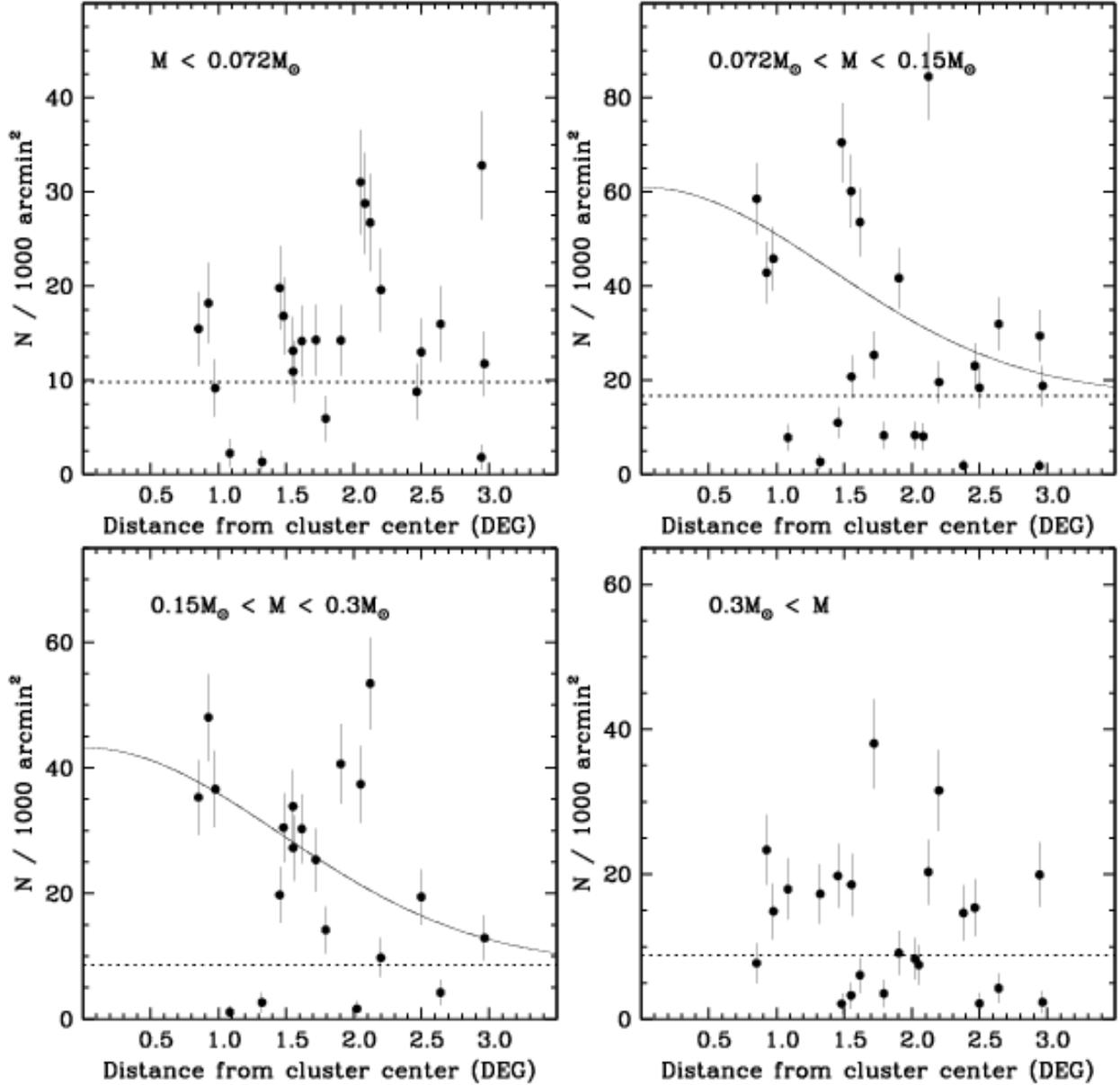}
  \caption{\label{fig:radial-prof} Radial profile of IC~2391 for four
    different mass bins as indicated in each panel. A King profile fit
    (\citealt{king1962}) was used for the radial profile for
    0.072\,M$_\odot$\,$<$,\,M\,$<$\,0.15\,M$_\odot$ and
    0.15\,M$_\odot$\,$<$,\,M\,$<$\,0.3\,M$_\odot$ (solid line). An
    estimation of the background contamination for each mass bin is
    given by the horizontal dotted line. In the panel of
    0.15\,M$_\odot$\,$<$,\,M\,$<$\,0.3\,M$_\odot$, the four data
    points at N\,=\,0 indicate that no object was detected in that
    mass range for these fields (fields 03, 04, 31, 40 and 41, where
    03 and 41 are at a similar distance of $\sim$2.94$^{\circ}$ from
    cluster center).}
\end{figure}

To help the analysis of the radial variation as a function of mass, we
also present on Figure \ref{fig:radial-prof} the radial profile of
IC~2391 using the radial fields for four different mass bins:
M\,$<$\,0.072\,M$_\odot$,
0.072\,M$_\odot$\,$<$\,M\,$<$\,0.15\,M$_\odot$,
0.15\,M$_\odot$\,$<$\,M\,$<$\,0.3\,M$_\odot$ and
0.3\,M$_\odot$\,$<$\,M.  (It is not surprising that a fit for the most
massive stars is not possible since the core radius is at 0.35$^\circ$
and we do not have any radial fields closer than 0.8$^\circ$.)  For
the second and third radial profile, we fit a King profile
(\citealt{king1962}), where the fit give us a maximal number density
at the center of 27.7$^{+29.9}_{-17.8}$ and 26.1$^{+16.9}_{-11.3}$
members per 1~000 arcmin$^2$ and a full width at half maximum of
1.39$^\circ$$^{+0.44}_{-0.23}$ (or 3.5\,pc$^{+1.2}_{-0.5}$) and
1.45$^\circ$$^{+0.47}_{-0.24}$ (or 3.7\,pc$^{+1.2}_{-0.6}$),
respectively. This is purely for illustration purposes, as it is clear
that this is not a good model for this mass profile.

In Figure \ref{fig:mf-r89j} we do not see any significant radial
variation of the MF in the substellar regime. The radial profile of
the substellar population in Figure \ref{fig:radial-prof} also
indicates no significant radial variation. On the other hand, we do
see reasonable evidence for a radial variation for masses above
0.072\,M$_\odot$. From this radial profile and the mass functions
already discussed, we can conclude that the spatial distribution of
the BD population is uniform compared to the stellar population (from
0.072 to 0.3\,M$_\odot$), which is more clustered within
$\theta$$\sim$2$^\circ$.

\cite{kumar2007} also found the stellar population to be more
clustered than the substellar population in the clusters IC~348 and
Trapezium.  This would favour the ejection scenario for forming BDs if
the BDs have a higher velocity dispersion than the stars
(\citealt{kroupa2003}), because the higher velocity from ejection
creates a more uniform spatial distribution for the BDs compared to
the stars.  However, the two clusters of \cite{kumar2007} are both
younger than 3\,Myr, while IC~2391 has an age of 50\,Myr, some $\sim$3
times older than its crossing time ($t_{cross}=$17\,Myr). We can
expect that if BDs have a higher velocity dispersion than stars, then
in an older cluster most of the BDs with velocity dispersion greater
than the escape velocity could have escaped the cluster already
(\citealt{morauxclarke2003}).

The homogeneous distribution of the substellar objects compared to the
more clustered stellar population could be instead a signature of mass
segregation through dynamical evolution or of primordial origin.  We
have indicated previously that mass segregation via dynamical
evolution could occur on a timescale of order one relaxation time
(\citealt{bonnell1998}), or even less (\citealt{allison2009}).
Considering that the cluster is only three times older than its
crossing time, which may be insufficient for significant dynamical
evolution, it is difficult to make an inference on the BD formation
mechanism from the radial MF variation given the uncertainty about to
what extend the cluster has evolved dynamically. If it could be
demonstrated that the cluster is dynamically unevolved and that the
BDs have a higher velocity dispersion than the stars, then our
observations are consistent with BD formation by the ejection
hypothesis.
 
\subsection{\label{contamination-rate} Contamination by non-members}

Possible point source contaminants other than field M--dwarfs include
red giants and high redshift quasars (\citealt{caballero2008}).
However, as pointed out in \S \ref{mf-deep-43-46-47-48-49} and in our
spectroscopic follow-up (see below in \S \ref{discussion-follow-up}),
the use of medium band filters and theoretical colours is efficient at
removing potential background red giant contaminants. As for the high
redshift quasars (for instance at $z$\,$\sim$\,6), their spectral
energy distribution is similar to mid-T dwarfs and moreover, they are
rare (\citealt{caballero2008}). Considering that our faintest targets
are early L--dwarfs, the MF should not be affected by contamination by
quasars.

Here we present an estimation of the contamination in our photometric
survey based on the radial fields. First, we used the radial profile
of IC~2391 using the radial fields for four different mass bin (Figure
\ref{fig:radial-prof}). We assumed that near the tidal radius
($\sim$2.89$^\circ$), the number of objects per 1~000\,arcmin$^2$
should be zero. From this, we computed that we can expect a
contamination of $\sim$8.8 objects per 1~000\,arcmin$^2$ for masses
above 0.3\,M$_\odot$, $\sim$8.8 objects per 1~000\,arcmin$^2$ in the
mass range of 0.15\,M$_\odot$\,$<$\,M\,$<$\,0.3\,M$_\odot$, $\sim$16.7
objects per 1~000\,arcmin$^2$ in the mass range of
0.072\,M$_\odot$\,$<$\,M\,$<$\,0.15\,M$_\odot$ and $\sim$9.8 objects
per 1~000\,arcmin$^2$ in the substellar regime. If we use the same
assumptions about the lognormal fit of the MF, the new total number of
object expected in IC~2391 is 1~954 for a total mass of
308\,M$_\odot$.

\section{\label{spec-follow-up} PRELIMINARY SPECTROSCOPIC FOLLOW-UP}

Here we present the results of a preliminary spectroscopic follow-up
of some photometric candidates. As explained in the previous section,
the main sources of contamination in our photometric selection are
background red giants and field M--dwarfs. We have also shown in \S
\ref{talk-ebv} that, because of extinction, background contamination
is non-uniform. We therefore need to refute or confirm membership
status with optical spectra. For this task we used the fiber
spectrograph HYDRA. It is not possible to cross fibres with this
instrument, so we have not yet been able to observe all candidates in
a given field.  (It is our intention to eventually obtain spectra of
all candidates.) The data reduction was described in \S \ref{spec}
while the spectral type and luminosity class determination was
presented in \S \ref{spt}.  The spectroscopic $T_{\rm eff}$ was
obtained using the spectral type and the temperature scales of
\cite{luhman1999b} while each mass was derived from $T_{\rm eff}$
using our isochrone for IC~2391.  We discuss membership determination
based on optical spectra below in \S \ref{member-spec}.

\begin{figure}
  \epsscale{1.0} \plotone{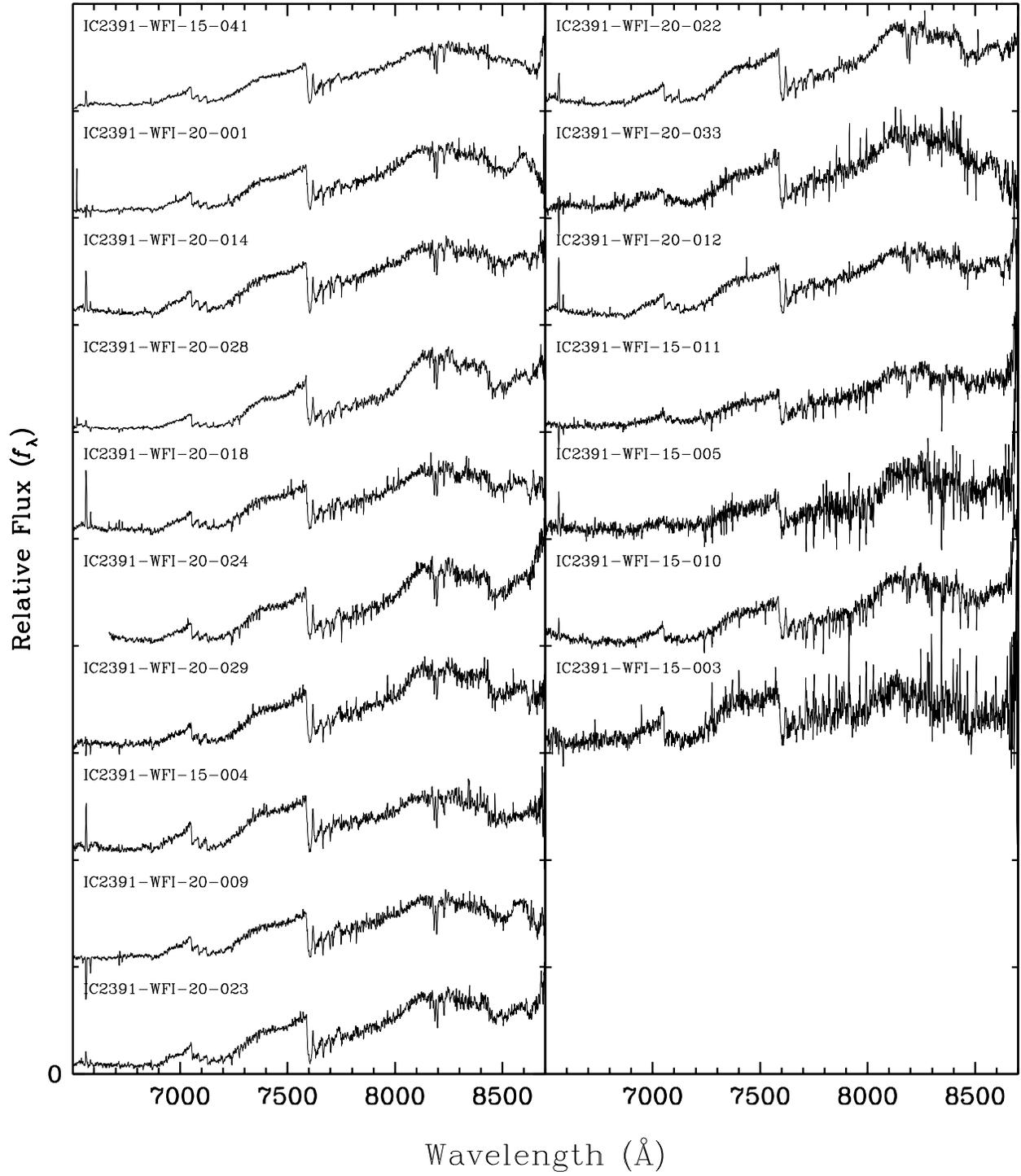}
  \caption{\label{fig:ic2391-20-1st} Spectra used in our analysis.}
\end{figure}

Among the spectra obtained, 17 had a signal-to-noise ratio (SNR)
higher than 5. These are presented in Figure \ref{fig:ic2391-20-1st}.
Table \ref{tab:allspec} provides the derived parameters (spectral
type, $T_{\rm eff}$ and mass) and the SNR.  Objects are given the same
notation as the photometric candidates : IC~2391-WFI-ZZ-YYY where ZZ
is the field number and YYY a serial identification number (ID).
Table \ref{tab:allspec2} gives details of an object confirmed as
cluster members by \cite{barrado2004}, including their SpT and $T_{\rm
  eff}$ determination, for which we also have a spectra.

\subsection{\label{h-alpha} Contaminating H$\alpha$ nebula emission}

We mentioned in \S \ref{spec} the presence of contamination at
H$\alpha$ in sky spectra. As these spectra are used for background
subtraction (we have fiber spectra), there are potential difficulties
in measuring the stellar H$\alpha$ line. We now discuss this issue.

\begin{figure}
  \epsscale{1.0} \plottwo{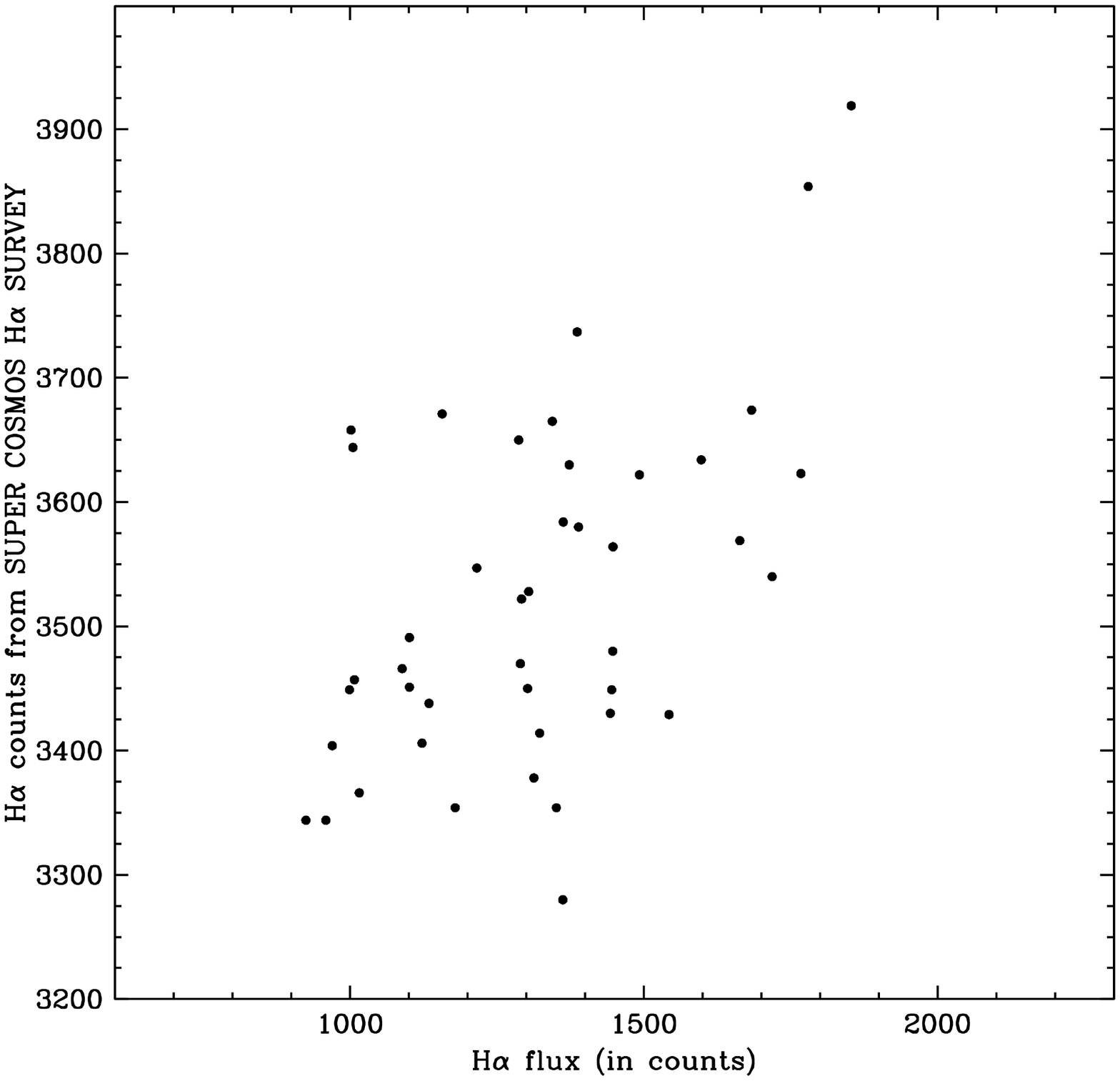}{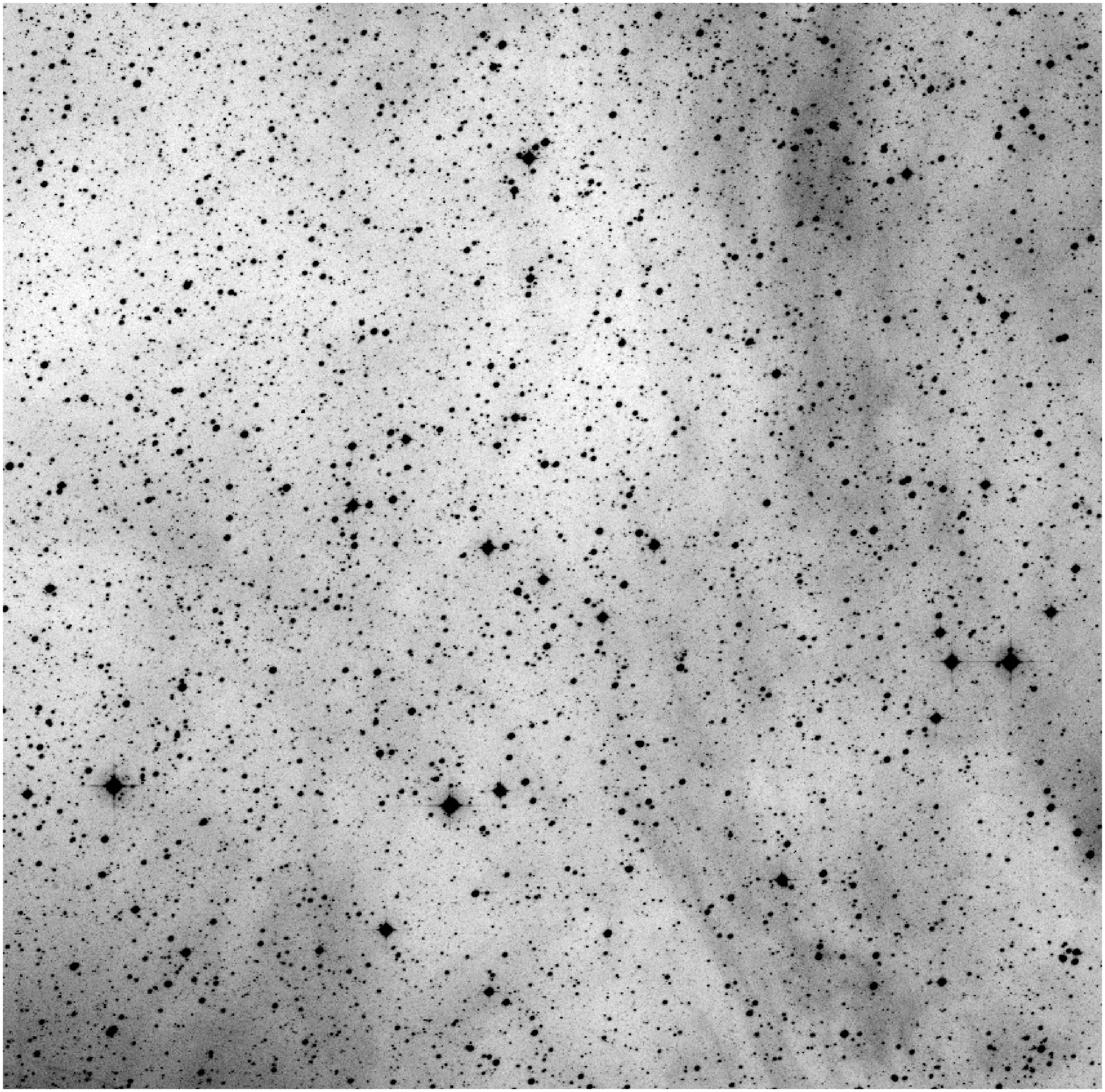}
  \caption{\label{fig:halpha_prob} \textit{Left.} H$\alpha$ emission
    (median counts) in fields of size 200 x 200 arcsec from the
    SuperComsos Survey at the location of our sky fibers, plotted
    against the flux (in counts) of the H$\alpha$ emission line we
    measured in our sky fibers in the HYDRA pointing of field 20.
    \textit{Right.} H$\alpha$ observations from SuperCOSMOS towards
    field 20 (images of 35$\times$35 arcmin).}
\end{figure}

In producing a high-resolution atlas of night-sky emission lines with
the Keck echelle spectrograph, \cite{osterbrock1996} observed an
H$\alpha$ emission line at high Galactic latitudes which they
concluded was due to diffuse interstellar gas emission (the closest
atmospheric emission observed were two OH lines at 6553.617\,\AA\, and
6568.779\,\AA\,).  From the AAO/UKST SuperCOSMOS H$\alpha$ Survey
(SHS, \citealt{parker2005}), we have also noticed high variations of
H$\alpha$ emission at low Galactic latitude. We used the SHS to
estimate the H$\alpha$ emission at each position of our sky fibers (by
taking a median of the flux over a 200 x 200 arcsec window).  The
frames are flat-field corrected but not flux calibrated, so we retain
the unit of (photon) counts.  In Figure \ref{fig:halpha_prob} we plot
this against the flux (in counts) of the H$\alpha$ emission line of
our background spectra for field 20. While there is no strong evidence
for a correlation, we nonetheless see a significant variation of the
H$\alpha$ emission. As this clearly prevents a reliable background
subtraction of the H$\alpha$ line, we choose not to draw any
conclusions on membership status based on this line.  We must
therefore question its use by \cite{barrado2004} for this purpose (who
used the same instrument for the same cluster).  For further
observations of objects in direction of IC~2391 using fiber-fed
spectrograph, we recommend background subtraction to be performed in a
similar way as the one done by \cite{carpenter1997}, where the same
fibers for the science targets were also used for sky subtraction but
shifted 6 arcsec away.

\subsection{\label{member-spec} Membership Determination}

We use the Li\,I line at 6708\,\AA\ to help confirm substellar status
of photometric candidates and to establish membership of IC~2391.
Lithium can be observed in young, more massive stars with radiative
interiors because of less efficient mixing than in fully convective
low mass stars (e.g.\ \citealt{manzi2008}).  Lithium may still even be
present in the atmospheres of young, fully convective low mass stars,
if they are young enough that not yet all the lithium has been
"burned" (\citealt{manzi2008}).  Older, lower mass BDs
($\lesssim$0.065\,M$_\odot$) never achieve core tempertures high
enough to burn lithium and so preserve their initial lithium content
(\citealt{rebolo1996}). Here we assume that field stars (with
M$\gtrsim$0.072\,M$_\odot$) are too old to still retain lithium in
their atmospheres. Hence we take the presence of the LiI line in
candidates with M$\lesssim$0.072\,M$_{\odot}$ as an indicator of
membership in IC~2391, as only cluster members fainter than the
Lithium depletion boundary ($I_{\rm c}=$16.2\,mag,
\citealt{barrado2001}) are young enough to retain Lithium.

The sodium doublet at 8200\,\AA\ is a gravity indicator, so is
sometimes used to exclude field stars. Specifically, its equivalent
width (EW) is sensitive to log\,$g$ (\citealt{martin1996}) and because
BDs contract as they age, log\,$g$ will increase.  Because field late
M--dwarfs (which have similar colours to M dwarf cluster members) will
generally be much older and so more evolved, they will have larger EWs
in this line (for a given chemical composition). We use the EW
measurement of CTIO-046 from the \cite{barrado2004} survey as a lower
limit on EW values for M-dwarfs to be non-members.  (This object was
defined as a non-member based on various criteria and had
$W$(NaI)=7.3$\pm$0.2\,\AA). It can be argued that, since surface
gravity changes with mass, the limit EW(NaI)=7.3\AA\ could also change
with mass.  Based on our isochrone of IC~2391, from 0.04 to
0.2\,M$_\odot$ (which is the mass range of our spectroscopic
follow-up), the surface gravity will varies only from log
\textit{g}\,=\,4.65 to 4.72 ($\delta$\,log \textit{g}\,=\,0.07).

The spectral resolution is sufficient to provide an estimate of the
radial velocity\footnote{The radial velocity measurment was performed
  with the IRAF task \textit{xcsao}. This task perform cross
  correlation against a spectrum with known radial velocity and makes
  the barycentric correction.}. As for the radial velocity criteria,
we exclude candidates which differ significantly ($\pm$3$\sigma$) from
a recent determination of the cluster's radial velocity
(16$\pm$3\,km/s, \citealt{kharchenko2005}, where $\sigma$ is the the
error of the radial veolicty of IC~2391 added in quadrature with the
error of our candidates). We didn't used radial velocity measurment
for which errors exceed 30\,km/s, which is ten time the error on the
radial velocity of IC~2391.

Finally, we use the SpT determination to obtain $T_{\rm eff}$ and
masses for each spectrum.  In order to be confirmed as a
(spectroscopic) cluster member, the spectroscopic $T_{\rm eff}$ and
must agree with the photometric $T_{\rm eff}$ to within 200\,K.

In Table \ref{tab:allspec3} we again present all objects from Table
\ref{tab:allspec}, but with physical parameters and with membership
status based on photometry and spectroscopy (i.e. which satisfy our
spectroscopic criterium). We don't reject objects below which does not
present a feature of LiI due to low SNR if the other criteria are
satisfied (e.g. IC2391-WFI-15-005).

\subsection{\label{discussion-follow-up} Discussion of the spectral data}

Of the 17 photometric candidates observed with a SNR higher than 5, 9
are spectroscopic members of the cluster. We find no red giants in our
spectral sample, which demonstrates that our choice of filters and
selection procedure is efficent at minimizing this contamination.
Since our spectroscopic follow-up covers only part of the mass range
used for the mass function calculation in \S \ref{mf-variation}, it is
not possible to compute a new MF with corrections applied at each mass
bin. It is expected that the contamination rate would be different for
other mass range and fields (for different filter combinations). One
of our spectral targets was observed by \cite{barrado2004}: CTIO-62,
which has the label IC2391-WFI-20-067 in our survey. We agree with
\cite{barrado2004} on the status (cluster member) of this object.

Another of our spectral targets, IC2391-WFI-20-001, shows Lithium even
though this object has an inferred mass above the stellar/substellar
boudary (spectroscopic mass of 0.079\,M$_\odot$ and a photometric mass
of 0.089\,M$_\odot$). The lithium depletion boundary has been
estimated to be at $I_{\rm c}=$16.2\,mag (\citealt{barrado2001}),
which corresponds to an effective temperature of about 3\,000\,K based
on our isochrones of IC~2391. As the cluster is not that old, the
lithium depletion boundary lies well above the substellar boundary, so
its not surprising to see a trace of Lithium in the spectrum of this
object (which has a magnitude of $I_{\rm c}=$16.681\,mag).

Although we obtained for IC2391-WFI-20-024 the same spectral type as
for IC2391-WFI-20-029, we consider this object as a member of IC~2391,
but to have a mass slightly above the substellar. Its photometry gives
$T_{\rm eff}$\,=\,2958\,K and M\,=\,0.081\,M$_\odot$, with a predicted
magnitude of 815/20\,=\,16.445\,mag.  If this object were an
unresolved binary, we would expect its observed magnitude to be
brighter than its predicted one, which is not the case (observed
magnitude of 815/20\,=\,16.520\,mag). Also, it should be noted that
this object was observed with a fibre with poor spectral response
below 6800\,\AA\, (which is why we don't show it in Figure
\ref{fig:ic2391-20-1st}).  Although this does not affect the PC3 index
used for the SpT determination (which covers 7540\,\AA\, to 7580\,\AA,
and 8230\,\AA\, to 8270\,\AA), it does influence the reduction process
(including the throughput correction, illumination correction,
extraction of the spectra and flux calibration).  Therefore, we
consider its SpT uncertainty to be larger (two subtypes rather than
one).

\subsection{\label{new-bd} Discovery of new brown dwarf members of
  IC~2391}

\begin{figure}
  \epsscale{1.0} \plotone{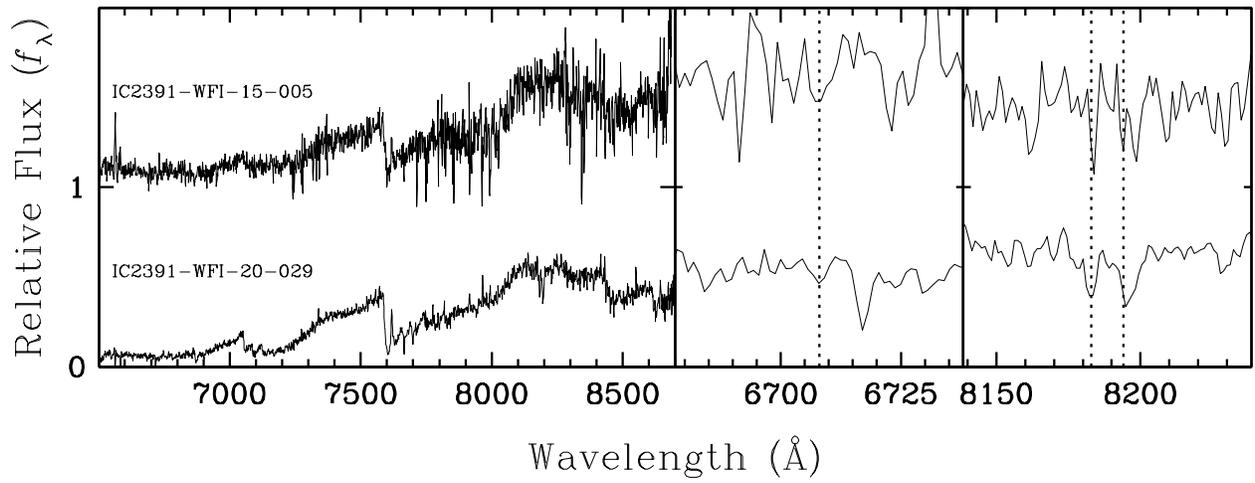}
  \caption{\label{fig:bd_4045} Spectra of the eight newly discovered
    brown dwarf members of the IC~2391 cluster found in our survey. We
    also present, for each spectra, a new close-up on the Li and the
    NaI doublet.}
\end{figure}

\begin{figure}
  \epsscale{1.0} \plotone{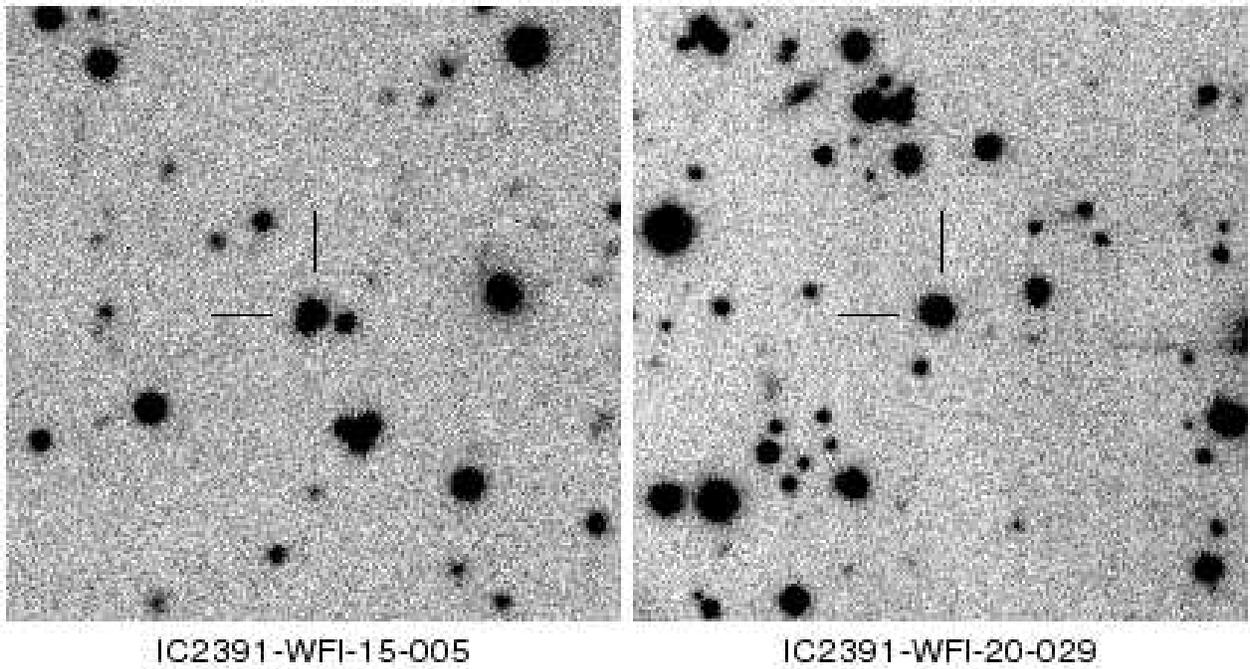}
  \caption{\label{fig:ic2391-bd} Finding charts of the two new brown
    dwarf members of IC~2391 (from 815/20 images). The panels are
    3.5$\times$3.5 arcmin with North up and East to the left.}
\end{figure}

Of the 17 spectral targets, we assign as brown dwarfs two in IC~2391,
on the basis of spectroscopic confirmation, and having both
photometric and spectroscopic masses below 0.072M\,$_\odot$. These are
new discoveries. These objects are IC2391-WFI-15-005 and
IC2391-WFI-20-029.  Table \ref{tab:allspecBD} lists their parameters,
Figure \ref{fig:bd_4045} shows their spectra and Figure
\ref{fig:ic2391-bd} contains the finding charts.  We can see in Figure
\ref{fig:bd_4045} that H$\alpha$ is not visible in IC2391-WFI-20-029,
so it would be designated as non-members by \cite{barrado2004}.
Considering that the MF from our radial fields is similar to that of
the deep fields in the mass range of these two new objects (from 0.045
to 0.07\,M$_\odot$), then if we used the same selection method, then
statistically we would expect to find two brown dwarfs in the same
mass range in the other two deep fields, and about 31 in all of the
radial fields.

\section{\label{conclusion} CONCLUSIONS}

We have performed a multi-band photometric survey over 10.9 square
degrees of the open cluster IC~2391, and completed a preliminary
spectroscopic follow-up of brown dwarfs and very low mass stars
candidates from two of the WFI fields.  Our objective was to study the
mass function of this cluster, and in particular its radial
dependence.  We observed a radial variation in the MF from 0.072 to
0.3\,M$_\odot$, but we do not observe a significant radial variation
in the mass function in the substellar regime. This comparative lack
of radial variation of the substellar mass function is in favour of
the ejection scenario for forming brown dwarfs, but considering that
IC~2391 is $\sim$3 times older than its crossing time, we might expect
that most of the brown dwarfs with velocity dispersion greater than
the escape velocity could have already escape the cluster. On the
other hand, the rather homogeneous distribution of the substellar
objects and the clustered distribution of stellar objects within
$\theta$$\sim$2$^\circ$ could be a signature that mass segregation via
dynamical evolution has occurred in IC~2391, or that this mass
segregation is of primordial nature. We have concluded that if this
cluster is dynamically unevolved and if the brown dwarfs have a higher
velocity dispersion than the stars, then our observations are
consistent with brown dwarf formation by the ejection hypothesis.

In addition to the radial study, we derived a mass function from four
central deeper fields as well as from five fields near the edge of the
cluster observed with only three filters (the outward fields).  In
both cases we see an apparent rise in the number of objects below
0.05\,M$_\odot$ (log$_{10}$M=-1.3), but we concluded that this is an
artefact of residual contamination by field M dwarfs. This was also
seen by \cite{barrado2004}.  The fact that we don't see this rise in
the radial fields is because they were observed with {\em both} the
$J$ and $R_{\rm c}$ filters in addition to the medium band filters.
This longer spectral baseline permits a better determination of the
energy distributions and thus helps the rejection of objects (in
particular field M dwarfs) based on observed magnitude vs.\ predicted
magnitude from models.

Another apparent rise in the MF over the 0.5--1.0\,M$_\odot$ interval
(also observed by \citealt{jeffries2004} for NGC~2547) is due to
background giants.  Red giant contamination may be reduced by using
medium bands such as 770/19, 815/20, 856/14 and 914/27, and
theoretical colours of red giants (\citealt{hauschildt1999b}). Our
spectroscopic follow-up has confirmed that selection based on these
filters resulted in no red giant contaminants among our sample of
spectra.

We see some variation in the colours of the main (field star) locus
which we attribute to variable extinction affecting the background
stars. This underlines the need for spectroscopic observations in this
cluster in order to confirm membership and/or brown dwarf status in
individual cases.

We have performed a preliminary spectroscopic follow-up of photometric
cadidates in two of our deep fields (0.5 sq.\ degrees). Of 17
photometric candidates, we confirm 9 objects (i.e.\ half) as true
cluster members.  Of these, two are new brown dwarf members of IC~2391
(in the sense that they fufill our spectroscopic and photometric
criteria).  Using our derived mass functions for the deep and radial
fields, we expect there to be two more brown dwarfs in the mass range
0.045 to 0.07\,M$_\odot$ in the other deep fields and up to 31 in all
the other radial fields in the same mass range.

Finally, we find that the H$\alpha$ line cannot be used as a
membership criterion from fiber spectroscopy at low spectral
resolution (spectral dispersion of 1.14\,\AA per pixel) because of
spatially variable diffuse H$\alpha$ emission. This prevents reliable
sky subtraction around this line when using a fiber spectrograph with
fibers assigned for sky subtraction.

\acknowledgments

S.B. and C.B.J. acknowledge support from the Deutsche
Forschungsgemeinschaft (DFG) grant BA2163 (Emmy-Noether Program) to
C.B.J. We are grateful to Reinhard Mundt for assistance with some of
the WFI observations. We also thank Matthew Coleman and Betrand
Goldman for their constructives comments on the paper. S.B. thanks the
CTIO mountain staff for support, in particular from Hern\'an Tirado.
Some of the observations on which this work is based were obtained
during ESO programmes 078.A-9056(A) and 079.A-9004(A) and NOAO
programmes 2006B-0251 and 2007A-0351. Some data analysis in this
article has made extensive use of the freely available R statistical
package, http://www.r-project.org. This research has made use of the
SIMBAD database, operated at CDS, Strasbourg, France.  This
publication makes use of data products from the \textit{Two Micron All
  Sky Survey}, which is a joint project of the University of
Massachusetts and the Infrared Processing and Analysis
Center$/$California Institute of Technology, funded by the National
Aeronautics and Space Administration and the National Science
Foundation. The DENIS project has been partly funded by the SCIENCE
and the HCM plans of the European Commission under grants CT920791 and
CT940627.

\clearpage



\clearpage

\clearpage

\clearpage

\clearpage

\clearpage

\clearpage

\clearpage

\clearpage

\clearpage

\clearpage

\clearpage

\clearpage

\clearpage

\clearpage

\clearpage

\clearpage

\clearpage

\clearpage

\clearpage

\clearpage

\begin{deluxetable}{ccccccccccc}
  \tabletypesize{\scriptsize} \rotate \tablecolumns{11}
  \tablewidth{0pt} \tablecaption{Description of observations.
    \label{tab:observations}} \tablehead{ \colhead{Field} &
    \colhead{RA} & \colhead{DEC} & \colhead{Distance ($^{\circ}$)} &
    \colhead{Region name} & \colhead{$R_{\rm c}$} & \colhead{770/19} &
    \colhead{815/20} & \colhead{856/14} & \colhead{914/27} &
    \colhead{$J$} } \startdata
  01&8:24:38.8&-51:18:16.5&2.966&radial&1500/23.5&-&1800/20.8&-&600/19.7&1820/17.8\\
  03&8:28:10.8&-50:46:50.0&2.945&radial&1500/23.8&-&1800/20.5&-&600/19.7&1820/17.7\\
  04&8:27:56.8&-52:07:47.9&2.084&radial&1500/23.9&-&1800/20.5&-&600/19.5&1820/17.5\\
  05&8:29:06.5&-52:26:14.6&1.793&radial&1500/23.7&-&1800/21.1&-&600/19.8&1820/17.5\\
  06&8:30:30.6&-51:38:59.1&2.051&radial&1500/23.8&-&1800/21.1&-&600/19.7&1820/18.0\\
  08&8:32:01.8&-52:15:26.1&1.481&radial&1500/23.9&-&1800/21.1&-&600/19.9&1820/17.9\\
  09&8:33:15.7&-50:39:24.6&2.640&radial&1500/23.9&-&1800/20.8&-&600/19.8&1820/18.0\\
  10&8:33:20.3&-51:50:14.6&1.616&radial&1500/24.0&-&1800/21.0&-&600/20.0&1820/18.0\\
  11&8:34:06.5&-51:24:35.9&1.904&radial&1500/23.5&-&1800/20.7&-&600/19.8&1820/17.9\\
  12&8:34:02.3&-52:45:43.2&0.978&radial&1500/23.7&-&1800/21.1&-&600/20.0&1820/17.9\\
  14&8:36:17.4&-50:38:44.9&2.498&radial&1500/23.4&-&1800/20.8&-&600/19.7&1820/17.9\\
  15&8:38:31.6&-53:35:29.4&0.757&deep&3900/22.7&3900/20.9&3000/20.7&1500/19.7&3000/20.3&-\\
  17&8:37:37.6&-51:34:05.6&1.552&radial&1500/23.6&-&1800/20.9&-&600/19.7&1820/17.5\\
  18&8:38:11.5&-52:01:44.1&1.085&radial&1500/22.7&-&1800/20.9&-&600/19.5&1820/17.6\\
  19&8:38:17.7&-50:58:08.4&2.121&radial&1500/24.0&-&1800/20.6&-&600/19.6&1820/17.7\\
  20&8:38:45.1&-52:35:58.0&0.519&deep&3900/22.4&3900/20.8&8400/21.3&15678/21.1&3000/20.3&-\\
  21&8:41:22.5&-52:14:04.7&0.854&radial&1500/23.8&-&1800/21.2&-&600/19.9&1820/17.7\\
  22&8:40:50.6&-51:31:11.6&1.553&radial&1500/23.7&-&1800/21.1&-&600/20.0&1820/17.8\\
  24&8:39:59.7&-54:14:24.0&1.170&radial&1500/23.8&-&1800/20.5&-&600/19.7&1820/17.9\\
  26&8:40:16.2&-55:16:12.0&2.200&radial&1500/23.7&-&1800/20.6&-&600/19.7&1820/17.7\\
  27&8:41:01.5&-53:50:51.7&0.789&deep&3900/22.5&3900/20.7&9300/21.5&7800/20.4&4800/20.7&-\\
  28&8:41:46.0&-54:46:27.8&1.720&radial&1500/23.8&-&1800/21.1&-&600/19.8&1820/17.5\\
  31&8:44:09.0&-55:27:58.0&2.464&radial&1500/23.6&-&1800/20.5&-&600/19.5&1820/17.5\\
  32&8:44:10.1&-52:39:49.0&0.721&deep&3900/22.3&3900/20.9&8450/21.4&10500/20.7&4800/20.5&-\\
  35&8:44:39.7&-54:21:53.7&1.453&radial&1500/23.5&-&1800/21.0&-&600/19.9&1820/17.2\\
  37&8:45:54.4&-53:25:53.5&0.927&radial&1500/23.5&-&1800/21.1&-&600/19.7&1820/17.5\\
  38&8:47:00.8&-53:55:12.1&1.323&radial&1500/22.5&-&1800/20.9&-&600/19.6&1820/17.7\\
  40&8:48:02.4&-55:09:16.4&2.380&radial&1500/22.8&-&1800/20.6&-&600/19.6&1820/17.7\\
  41&8:48:09.8&-55:46:27.3&2.942&radial&1500/22.7&-&1800/20.4&-&600/19.4&1820/17.7\\
  42&8:49:10.6&-54:35:58.2&2.023&radial&1500/22.7&-&1800/21.1&-&600/19.9&1820/17.5\\
  43&8:50:15.8&-53:23:37.1&1.540&outward&-&-&1800/20.5&-&600/19.3&1820/17.5\\
  46&8:53:03.1&-54:23:52.2&2.318&outward&-&-&1800/20.5&-&600/19.6&1820/17.3\\
  47&8:53:26.1&-53:52:29.4&2.127&outward&-&-&1800/20.9&-&300/19.3&1820/17.3\\
  48&8:54:39.8&-53:29:44.4&2.203&outward&-&-&1800/20.5&-&600/19.7&1820/18.2\\
  49&8:56:47.5&-54:17:22.5&2.742&outward&-&-&1800/20.7&-&600/19.6&1820/17.8\\
  \enddata 
  \tablecomments{System notation is \textit{exposure time in seconds}
    / \textit{10$\sigma$ detection limit} while -- indicate that no
    observations was performed for that field in that filter.
    \textit{Distance} give the distance of the field from cluster
    center (in degree).  The 10$\sigma$ detection limit in $R_{\rm c}$
    is smaller than for the deep fields, although the exposure time
    was higher. $R_{\rm c}$ observations of the deep field were done
    in January 2000, while the observations of the radial fields were
    done in April 2007. Condensation problems and non-photometric
    nights was reported for the January 2000 observation run.}
\end{deluxetable}

\clearpage
 
\begin{deluxetable}{ccccccccccccc}
  \tabletypesize{\scriptsize} \rotate \tablecolumns{13}
  \tablewidth{0pt} \tablecaption{All photometric candidates of our
    survey\label{tab:allphot}} \tablehead{ \colhead{Field} &
    \colhead{ID} & \colhead{RA} & \colhead{DEC} & \colhead{$R_{\rm c}$} &
    \colhead{770/19} & \colhead{815/20} & \colhead{856/14} &
    \colhead{914/27} & \colhead{$J$} & \colhead{M} & \colhead{$T_{\rm
        eff}$} & \colhead{[815/20]} } \startdata
  01& 001&8:26:16.055&-51:02:52.32&19.645&$-$&16.706&$-$&16.292&15.023&0.057&2768&17.220\\
  01& 002&8:26:14.211&-51:02:47.64&18.426&$-$&16.200&$-$&15.920&15.145&0.104&3072&15.943\\
  01& 003&8:25:28.826&-51:13:56.71&18.866&$-$&16.383&$-$&15.998&14.909&0.081&2957&16.450\\
  01& 004&8:25:47.698&-51:06:25.26&18.848&$-$&16.332&$-$&15.953&14.969&0.078&2935&16.537\\
  01& 005&8:25:26.432&-51:03:59.27&18.359&$-$&16.077&$-$&15.771&14.962&0.096&3043&16.080\\
  01& 006&8:25:01.148&-51:03:37.82&19.574&$-$&16.953&$-$&16.582&15.446&0.072&2890&16.710\\
  01& 007&8:24:10.028&-51:07:26.14&19.283&$-$&16.735&$-$&16.341&15.427&0.076&2925&16.576\\
  01& 008&8:24:33.508&-51:05:35.22&18.764&$-$&16.437&$-$&16.094&15.130&0.093&3029&16.146\\
  01& 009&8:23:19.415&-51:32:35.82&19.900&$-$&17.106&$-$&16.655&15.618&0.065&2838&16.930\\
  01& 010&8:23:29.455&-51:32:00.48&18.373&$-$&16.076&$-$&15.762&15.014&0.094&3032&16.134\\
  \enddata
  \tablecomments{Table \ref{tab:allphot} is published in its entirety
    in the electronic edition of the {\it Astrophysical Journal}.  A
    portion is shown here for guidance regarding its form and content.
    The error on the determination of masses and effective temperature
    are the following~: $\Delta$$T_{\rm eff}$\,=\,140\,K and
    $\Delta$M\,=\,0.1\,M$_\odot$ for stars (M\,$>$\,0.2\,M$_\odot$),
    $\Delta$$T_{\rm eff}$\,=\,230\,K and $\Delta$M\,=\,0.05\,M$_\odot$
    for VLMS (0.072\,$_\odot$\,$<$\,M\,$<$\,0.2\,M$_\odot$),
    $\Delta$$T_{\rm eff}$\,=\,420\,K and $\Delta$M\,=\,0.02\,M$_\odot$
    for BDs (M\,$<$\,0.072\,M$_\odot$). The magnitude [815/20] is the
    predicted magnitude based on photometric determination of $T_{\rm
      eff}$ and mass.}
\end{deluxetable}

\clearpage

\begin{deluxetable}{cccccccccccc}
  \tabletypesize{\scriptsize} \rotate \tablecolumns{12}
  \tablewidth{0pt} \tablecaption{Objects from \cite{barrado2004},
    \cite{dodd2004} and detected by XMM-Newton which are photometric
    candidates in our sample \label{tab:allphot2}} \tablehead{
    \colhead{Field} & \colhead{ID} & \colhead{RA} & \colhead{DEC} &
    \colhead{815/20} & \colhead{M} & \colhead{$T_{\rm eff}$} &
    \colhead{[815/20]} & \colhead{NAME} & \colhead{$Ic$} &
    \colhead{($R-I$)$c$} & \colhead{$T_{\rm eff}$} } \startdata
  18&006&8:38:47.074&-52:14:56.16&17.076&0.053&2723&17.396&CTIO-061&17.309&2.141&2801\\
  20&028&8:38:47.282&-52:44:32.61&16.662&0.077&2927&16.567&CTIO-062&16.765&2.000&2937\\
  27&002&8:40:09.537&-53:37:49.81&16.153&0.095&3036&16.115&CTIO-077&16.308&1.929&2960\\
  32&120&8:44:02.109&-52:44:10.73&17.050&0.065&2842&16.911&CTIO-160&17.151&2.090&2806\\
  32&231&8:43:38.421&-52:50:55.15&14.810&0.206&3310&14.707&8       &14.890&1.790& -  \\
  32&295&8:43:38.422&-52:50:55.13&14.848&0.206&3310&14.707&CTIO-152&14.891&1.781&3053\\
  32&295&8:43:38.422&-52:50:55.13&14.848&0.206&3310&14.707&155     &14.530&2.260& -  \\
  32&325&8:46:15.404&-52:49:37.61&15.305&0.170&3249&15.046&2XMM~J084615.3-524937&   -  &  -  & -  \\
  32&340&8:46:04.238&-52:45:18.99&15.921&0.122&3134&15.640&2XMM~J084604.3-524518&   -  &  -  & -  \\
  37&024&8:47:07.572&-53:09:45.32&15.377&0.159&3228&15.164&2XMM~J084706.2-530944&   -  &  -  & -  \\
\enddata
\tablecomments{Object 8 is from \cite{patten1999}, CTIO objects are
  from \cite{barrado2004}, object 155 is from \cite{dodd2004} while
  the 2XMM objects are from XMM-Newton. The values of $Ic$, ($R-I$)$c$
  and $T_{\rm eff}$ given in the last three columns are taken from
  each previous studies.}
\end{deluxetable}

\clearpage

\begin{deluxetable}{cccccccccccc}
  \tabletypesize{\scriptsize} \rotate \tablecolumns{12}
  \tablewidth{0pt} \tablecaption{Stellar parameters from the objects
    observed in our spectroscopic follow-up. \label{tab:allspec}}
  \tablehead{ \colhead{Field} & \colhead{ID} & \colhead{RA} &
    \colhead{DEC} & \colhead{SNR} & \colhead{SpT} & \colhead{$T_{\rm
        eff}$} & \colhead{M} & \colhead{815/20} & \colhead{$T_{\rm
        eff}$ (phot)} & \colhead{M (phot)} & \colhead{[815/20]} }
  \startdata
  15&003&8:36:40.363&-53:21:30.00& 6.1&M2.0&3510 &0.413&16.201&3020 &0.091&16.191\\
  15&004&8:36:04.075&-53:29:25.00&20.7&M3.5&3265 &0.173&16.306&3010 &0.089&16.232\\
  15&005&8:36:18.241&-53:25:57.60& 8.3&M7.5&2660 &0.049&17.803&2654 &0.048&17.672\\
  15&010&8:35:21.606&-53:42:05.12& 8.0&M7.5&2660 &0.049&16.473&2955 &0.081&16.456\\
  15&011&8:38:06.120&-53:38:10.55& 9.0&M7.0&2720 &0.050&16.651&2925 &0.076&16.576\\
  15&041&8:36:54.383&-53:45:42.10&35.2&M5.5&2925 &0.073&14.988&3266 &0.179&14.953\\
  20&001&8:40:34.407&-52:30:38.65&32.0&M5.0&3010 &0.089&16.681&2947 &0.079&16.489\\
  20&009&8:38:36.649&-52:27:47.16&20.2&M5.0&3010 &0.089&16.873&2911 &0.075&16.631\\
  20&012&8:37:38.539&-52:29:35.37& 9.2&M4.5&3095 &0.110&16.645&2956 &0.081&16.453\\
  20&014&8:37:34.755&-52:27:02.90&27.6&M5.0&3010 &0.089&16.307&3026 &0.093&16.161\\
  20&018&8:37:58.411&-52:20:30.69&22.0&M5.0&3010 &0.089&16.907&2890 &0.072&16.710\\
  20&022&8:38:47.282&-52:44:32.61&17.0&M6.0&2840 &0.065&16.662&2927 &0.077&16.567\\
  20&023&8:39:22.724&-52:50:34.42&20.0&M2.0&3510 &0.413&16.433&2998 &0.087&16.280\\
  20&024&8:39:33.260&-52:47:10.40&21.8&M6.5&2780 &0.056&16.520&2958 &0.081&16.445\\
  20&028&8:40:15.153&-52:40:24.56&26.5&M7.5&2660 &0.049&16.494&2955 &0.080&16.459\\
  20&029&8:40:16.671&-52:36:58.32&21.4&M6.5&2780 &0.056&17.571&2740 &0.055&17.329\\
  20&033&8:39:27.836&-52:32:58.62&13.1&M6.5&2780 &0.056&14.856&3297 &0.197&14.783\\
  \enddata
  \tablecomments{The error on the determination of masses and
    effective temperature based on spectroscopy are the following~:
    $\Delta$$T_{\rm eff}$\,=\,190\,K and $\Delta$M\,=\,0.03\,M$_\odot$
    for spectra with SNR\,$>$\,10, $\Delta$$T_{\rm eff}$\,=\,320\,K
    and $\Delta$M\,=\,0.04\,M$_\odot$ for spectra with
    5\,$<$\,SNR\,$<$\,10. This would corresponds to an error on the
    spectral determination of 1 and 1.5 for spectra with SNR\,$>$\,10
    and 5\,$<$\,SNR\,$<$\,10 respectively.}
\end{deluxetable}

\clearpage

\begin{deluxetable}{ccccccccccc}
  \tabletypesize{\scriptsize} \rotate \tablecolumns{11}
  \tablewidth{0pt} \tablecaption{Object from \cite{barrado2004} which
    we also observed in our spectroscopic follow-up.
    \label{tab:allspec2}} \tablehead{ \colhead{Field} & \colhead{ID} &
    \colhead{RA} & \colhead{DEC} & \colhead{SNR} & \colhead{SpT} &
    \colhead{$T_{\rm eff}$} & \colhead{M} & \colhead{NAME} &
    \colhead{SpT} & \colhead{$T_{\rm eff}$} } \startdata
  20&022&8:38:47.282&-52:44:32.61&17.0&M6.0&2840 &0.065&CTIO-062&M6.0&2800\\
\enddata
\tablecomments{The SpT and $T_{\rm eff}$ given in the last two columns
  are based on the spectra from \cite{barrado2004}.}
\end{deluxetable}

\begin{deluxetable}{cccccccccc}
  \tabletypesize{\scriptsize} \rotate \tablecolumns{10}
  \tablewidth{0pt} \tablecaption{\label{tab:allspec3} Spectroscopic
    data, photometric and spectroscopic membership status} \tablehead{
    \colhead{Field} & \colhead{ID} & \colhead{SNR} & \colhead{SpT} &
    \colhead{$EW$(H$\alpha$)} & \colhead{$EW$(NaI 8182\,\AA)} &
    \colhead{$EW$(NaI 8194\,\AA)} & \colhead{RV (km/s)} &
    \colhead{$EW$(LiI)} & \colhead{Spec member ?}  } \startdata
  15&003& 6.1& M2.0& -   & 1.21& -   & -    & - & NO\\
  15&004&20.7& M3.5& 7.38& 3.12& 2.43& -    & - & NO\\
  15&005& 8.3& M7.5& 4.30& -   & -   & -    & - &YES\\
  15&010& 8.0& M7.5& 2.80& 0.75& -   & -    & - & NO\\
  15&011& 9.0& M7.0&-6.80& 2.15& 1.77& -    & - & NO\\
  15&041&35.2& M5.5&12.66& 1.02& 1.75& -    & - & NO\\
  20&001&32.0& M5.0& -   & 1.21& 2.88& 15.11( 8.39 )& 0.8 &YES\\
  20&009&20.2& M5.0&-17.56& 2.65& 2.19&-18.28(12.37 )& $<$0.1 &YES\\
  20&012& 9.2& M4.5&15.18& -   & -   & 11.31(27.70 )& - &YES\\
  20&014&27.6& M5.0&13.24& -   & -   & 23.33( 6.88 )& $<$0.1 &YES\\
  20&018&22.0& M5.0&21.36& 1.91& 3.32& 19.32( 7.78 )& - &YES\\
  20&022&17.0& M6.0&13.03& 1.55& 2.61& 14.95( 8.14 )& - &YES\\
  20&023&20.0& M2.0&-6.12& -   & 1.72& 94.74( 6.90 )& - & NO\\
  20&024&21.8& M6.5& -   & 1.29& -   & 27.91( 7.35 )& - &YES\\
  20&028&26.5& M7.5& -   & 1.23& -   & 11.78( 7.48 )& - & NO\\
  20&029&21.4& M6.5& -   & 2.38& 2.33& 16.25( 6.59 )& 1.3 &YES\\
  20&033&13.1& M6.5& 5.80& -   & 1.96&  6.23( 7.04 )& - & NO\\
\enddata
\tablecomments{A value of \textit{$<$0.1} as equivalent width of
  Lithium indicates the presence of a feature at 6708\,\AA\, but no
  reliable measurement of equivalent width can be done.}
\end{deluxetable}

\begin{deluxetable}{cccccccccccc}
  \tabletypesize{\scriptsize} \rotate \tablecolumns{12}
  \tablewidth{0pt} \tablecaption{Stellar parameters obtained from
    spectra. \label{tab:allspecBD}} \tablehead{ \colhead{Field} &
    \colhead{ID} & \colhead{RA} & \colhead{DEC} & \colhead{SNR} &
    \colhead{$R_{\rm c}$} & \colhead{$R_{\rm c}$-815/20} & \colhead{SpT} &
    \colhead{$T_{\rm eff}$} & \colhead{M} & \colhead{$T_{\rm eff}$
      (phot)} & \colhead{M} (phot) } \startdata
  15&005&8:36:18.241&-53:25:57.60& 8.3&21.048& 3.245&M7.5&2660&0.049&2654&0.048\\
  20&029&8:40:16.671&-52:36:58.32&21.4&20.076& 2.505&M6.5&2780&0.056&2740&0.055\\
\enddata
\end{deluxetable}







\end{document}